\documentclass{article}
\usepackage[utf8]{inputenc}
\usepackage{amssymb, amsmath, amsthm, amsfonts}
\usepackage[onehalfspacing]{setspace}
\usepackage{graphicx}
\usepackage[english]{babel}
\usepackage{tikz}
\usepackage{caption}
\usepackage{subcaption}
\usepackage{float}
\usepackage{natbib}
\newtheorem{theorem}{Theorem}

\title{Moment-Based Estimation of Diffusion and Adoption Parameters in Networks 
}

\author{L.S. Sanna Stephan
\footnote{I sincerely thank Prof. Matthew O. Jackson from Stanford University, whose ideas and suggestions have been essential for the development of the proposed estimator.}
}
\begin{document}

\maketitle

\section{Introduction}

According to standard econometric theory, Maximum Likelihood estimation (MLE) is the efficient estimation choice, however, it is not always a feasible one. In network diffusion models with unobserved signal propagation, MLE requires integrating out a large number of latent variables, which quickly becomes computationally infeasible even for moderate network sizes and time horizons. Limiting the model time horizon on the other hand entails loss of important information while approximation techniques entail a (small) error that. Searching for a viable alternative is thus potentially highly beneficial. This paper proposes two estimators specifically tailored to the network diffusion model of partially observed adoption and unobserved network diffusion.  \\ \noindent 
Akin to Generalized Method of Moments (GMM) estimators, they are constructed using moment conditions and minimize the deviation of individual outcomes from their unconditional expected value. They distinguish themselves from usual GMM estimators in three important ways: first, moments are specific to individuals. This also implies that the correlations in individual moment-conditions are pair-specific. Second, these correlations  potentially entail small sample bias when individual moment conditions are aggregated. Third, for computational reasons outlined below, each agent will contribute only to one moment, resulting in moment-specific sample sizes whenever one chooses to aggregate individual moment conditions.      
\\ \noindent 
Moment-based estimators have previously been derived for and applied to social network models. 
Two strands of research can be distinguished. 
First, various authors have applied moment-based estimates to network formation models where the aim is to uncover the unobserved network structure (\citet{NWformation1}, \citet{NWformation2}). 
Second, these techniques have been used for fitting models of observed network interaction in general (\citet{NW1}) and spatial-auto-regressive models in particular (\citet{NW2}).
However, to the best of my knowledge, moment-based estimation has not been applied to the ``latent-diffusion-observed-adoption" model of \citet{BCDJ}.
Given the widespread applicability of the latter, the contribution made in this paper is thus important and relevant. 
\\ \noindent
Network models entail the challenge that observations for individuals within the same network are in general correlated. Correlation between the individual observations that are aggregated into a moment is a challenge typically encountered in Time Series analysis. Various papers have shown that the GMM estimator is consistent if the data is mean stationary and ergodic (see \citet{TS1}). The main difference between the application of GMM estimation to Time Series and Network Models is that in the latter case, while correlations are typically bounded and decay to zero asymptotically, yet no such notion as covariance stationarity exists, making the estimation of the estimators' covariance matrices computationally more challenging. \citet{asycluster} have derived conditions for asymptotic normality and consistent covariance estimation in clustered samples, a theory that applies to network models. Yet it is challenging at this point to state how covariance estimates based on this theory perform in finite samples. \\ \noindent
In the model at hand, as individuals exchange information in the network, the network inter dependencies solely concern the information reception probabilities, which are hard to evaluate due to the large number of possibilities by which an agent may have been reached by the information.
Since any moment requires computation of these probabilities, moment-based estimation in principle still suffers from the same problems as MLE. Calculating the probability of any particular agent to receive the information at any point in time $t$ requires integrating out the entire information status matrix of all villagers from period one to period $t-1$.  
The remedy of the dimensionality problem is not so much the idea to use moments itself, but the fact that a moment-based approach enables me to focus on the set of moment conditions that are easy to evaluate. 
As the system is over-identified, no approximation technique is needed, exact estimation is feasible. 
Precisely, an agent's information reception probability can straightforwardly be computed in the very first period in which she can be reached by the information. Here, some shorthand formulas exist that do not require integrating out the neighbours' information scenarios. These formulas are correct in the one and only period which is the first in which an agent can be reached by the signal. 
What makes rates hard to compute in subsequent periods is the fact that the information may stop or travel on when it reaches intermediate agents and that the latter can thereafter exchange the information among themselves. 
This increases the number of possible scenarios. 
When we consider an agent's first information reception opportunity, on the other hand, the information can only possibly have travelled directly from its ``injection point" (IP) to the agent.
Therefore, only the direct path from the agent to (an) IP(s) matter(s) and linkages between intermediate agents on that path can be neglected.
As a consequence, the number of scenarios that are needed to be considered in order to compute the (first-opportunity) reception probability are kept small and easily tractable formulas can be used for most agents. 
Notably, these formulas do get more and more sophisticated for agents further away from the information source, but for an intermediate model time horizon (here $T=4$) they remain tractable. 
\\ \noindent Due to the network dependencies, individual moments are in general correlated within the village. 
If the sample size increases while village sizes are kept constant (e.g.\ the increase in the sample size is achieved by adding more villages), the variance of the estimator does converge to zero, albeit at a rate smaller than the one observed for usual GMM estimators. This is intuitive: if individuals contribute to only one moment condition, adding more individuals will not decrease the variance proportionally. %The finite sample variance of the estimator can be approximated if the absence of correlation between villages dominates over the presence of correlation within villages.   
 \\ \noindent 
% \\ \noindent Yet this strategy can be challenging in finite samples. If the number of initially informed individuals is chosen too low, then few individuals dispose over direct or indirect IP-connections. If there are too many IPs on the other hand, then there are no or very few individuals fulfilling the conditions of \ref{N3} such that the last round of information diffusion and participation gets completely neglected. \\ \noindentWith few individuals that can be used in the moment conditions, the variance of the resulting estimator will be large. How many usable observations we expect to obtain depends on both the (expected)  network and the number of IPs. Presuming that the network is formed according to the Erdós-Renyi network formation model, I investigate how varying the number of IPs (as a fraction of the network size $N$) translates into the parameter variance for different values of link formation probabilities.  \\ \noindentEstimation in a concrete realized sample is tremendously simplified when the objective function is (approximately) convex. In this case, a standard Newton-Raphson algorithm can be applied and the estimate converges to the unique peak irrespective  of the starting point. Given a concrete realized network, information initiation and participation pattern, I analyze the convexity conditions, pointing out which features of the data may lead to a non-convex objective function. 
The remainder is organized as follows. First, the model and assumptions are recapitulated. 
Individual-specific moment conditions are the basis for any moment-based estimation in this model.
Section 3 introduces these conditions and explains how they can be calculated and used for parameter identification. This section includes the main novelty of the estimation approach, namely the insight that individual means are in general hard, but under certain conditions easy to calculate. It also discusses the challenges one faces when wanting to set up an estimator based on the individual-specific moment conditions. The following two sections propose two such estimators and analyses their properties. Section 6 provides a comparison, listing similarities and differences of the two estimators. Section 7 presents evidence from a Monte Carlo study that compares the estimates and section 8 a small application of the estimators to the real data. Section 9 concludes. Proofs of the relevant theories are provided in the Appendix.  

\section{The Model}

In this and the following sections, I use the terminology of the concrete application despite the fact that the model and the estimation technique are applicable to many research questions. \\ \noindent
We dispose over a sample of villages, indexed by $v=1,...,V$. 
\\ \noindent $N_v$ individuals in village $v$ are linked in a network, represented by the network adjacency matrix $G_v$. While villagers $i=1,...,N_v$ entertain  relationships among themselves, they are supposed not to interfere with agents living in other villages. The symmetric matrix $G_v$ comprises binary variables indicating the presence (absence) of a link. 
If the entry in row $i$, column $j$ takes the value of one (i.e. $g_{v,ij}=g_{v,ji}=1$), we denominate $i$ and $j$ as “neighbors" or “friends". An organization enters the community and starts providing a new technology to its members. 
In period $t=0$, they advertise their innovation  to a subset of them (referred to as the “information injection points" or “IPs") and  thereafter rely on word-of-mouth marketing.   \\
\noindent
In each subsequent period ($t=1,...,4$), two processes take place: first, newly informed individuals face the choice of whether or not to adapt the technology, second, informed individuals can instruct their neighbors about the novel opportunity. \\ \noindent 
The process is modeled using the random matrices $Y_v$ and $S_v$, containing dummy variables for respectively the participation and information statuses of each of the villagers at each point in time. Let $Y$ and $S$ refer to the sample outcome and information status matrices, respectively.  
Let $Y_{it}=1$ ($S_{it}=1$) (i.e.\ the row $i$, column $t$ element of $Y$ ($S$) displays the value one) implies that $i$ participates (is informed) in period $t$. 
The distributions of the random matrices $Y$ and $S$ depend on $G_v$ and $s_{v,0}$. %(see Appendix 1). 
\\
\noindent
For each village, we observe the network ($G_v$), the information initialization $s_{v,0}$.
and the individual participation decisions over time (i.e.\ one realization of $Y_v$). The information statuses of all inhabitants but the IPs are generally unobserved. In particular, there usually exist various realizations of $S_v$ that are in accordance with the data at hand. \\
\noindent The following assumptions are made: \\

\indent \textbf{Assumption 1:} Independence across villages. \\
Villages can be treated as independent entities. In particular, there is no link between any two agents that reside in different communities. \\
\\
\indent \textbf{Assumption 2:} Exogenous network. \\
The network is exogenous, fixed and observed. Measurement error is negligible. 
\\ \\
\indent \textbf{Assumption 3:} Timing. \\
Each period consists of two processes: First, newly informed individuals decide upon participation, second, information is exchanged.
This implies that $Y$ is of dimension $N_v \times 4$ while $S$ is of dimension $N_v \times 3$: since information is exchanged after the participation decisions, modeling the last period's information exchange is redundant.
\\ \\
\indent \textbf{Assumption 4:} Information is a pre-condition for participation. \\ \\
\indent \textbf{Assumption 5:} Participation is a one-time opportunity. \\ \noindent
Each period only \underline{newly informed} individuals face the participation decision. Having opted in (out), the respective individual will thereafter stay in the set of participants (non-participants) forever. \\
\\
\indent \textbf{Assumption 6:} Distributional assumption. \\
Conditional on being newly informed, the random variables $ Y_{it}
\hspace{0,15cm} \forall i=1,...,N_v; \hspace{0,15cm} t=1,...,4$ are i.i.d. and  follow a Bernoulli distribution. The probability to participate conditional on being newly informed is $p$. \\ \\
\indent \textbf{Assumption 7:} Information is never forgotten. \\
Individuals can only switch their information status once (from being uninformed ($S_{i(t-1)}=0$) to being informed ($S_{it}=1$)). Any informed individual remains potentially informing her neighbors every period.  
\\ \\
\indent \textbf{Assumption 8:} Information exchange. \\
In each period, informed individuals may transmit the information to their neighbors. On average, they do so with probability $q$. Transmitting is independent across individuals.  \\ \noindent
Let the random variable $I_{ij,t}$ denote the indicator  that shows that individual $i$ has sent the information to individual $j$ in period $t$. 
Conditional on $i$ being a sender, the variables $I_{ij,t} \hspace{0,1cm} \forall i\neq j; t=1,...,3 $ are i.i.d.\ and follow a Bernoulli distribution with $E[I_{ij,t}]=q$. 
\\ \\ \noindent
The aim is to estimate \\
\indent (i) the individual's probability to participate if informed ($p$) and \\
\indent (ii) individual's probability to share information with acquaintances ($q$)

\section{Unconditional Individual-Specific Moments and Moment Conditions}

\subsection{Individual-specific Mean-Conditions}

From the above it is apparent that the outcomes follow a Bernoulli distribution with individual and time specific means. Let 
\[ \mu_{it} (p,q,S_0,G)=E[Y_{it}|p,q,S_0,G] \]
be the unconditional expected value of individual $i's$ outcome in period $t$. As the Bernoulli distribution is characterised in terms of its mean, it is indispensable for any moment-based estimation to first derive a closed-form expression for $\mu_{it}$. 
 The data comprises $N \times T$ outcomes, for which $N \times T$ means can be derived. The mean depends on whether and how the individual is connected to the information and hence it is a function of the network, the information initiation and the parameters and although its concrete functional form potentially differs across agents, it is computed using the same (time period specific) formulas (outlined below) for all agents.
 Once an expression for $\mu_{it}$ is available, it can be used to construct individual-specific moment conditions that can be used to set up a criterion function for estimation. The most simplistic moment-condition is obtained by taking the individual's outcome and subtracting its mean, e.g. 
\[ g_{it}=Y_{it}-\mu_{it}(p,q,S_0,G)  \]
Clearly, 
\[ E[g_{it}]=0 \]
\noindent
Due to network dependencies, the attempt to calculate $\mu_{it}$ can lead to prohibitively complicated expressions. 
However, for the period that follows the first period in which an agent can be reached by the information, the unconditional mean can be straightforwardly derived. This is the aim of this section. 
 \\
\noindent
As stated, for each agent, I wish to solely use the first period in which the agent potentially faces a choice, e.g.\ the first period which they may enter being informed. In the following, I use $N$ ($N_v$) to denote the number of individuals in the sample (in village $v$) that can be reached by the information within the modelled time horizon.
Let $\tilde{N}_t $ be a vector of indicator variables used to identify the set of individuals who enter the information radius at time $t-1$ and who fulfill certain conditions outlined below (e.g.\ $\tilde{N}_{it}=1$ implies that period $t-1$ is the first period in which individual $i$ has a positive probability to be informed). 
Note that these indicators solely depend on the network and are thus observable. 
Being in reach at time $t-1$ implies that the individual faces a choice for the first time at time $t$. Exact estimation is feasible if only the set of individuals for which $\tilde{N}_{it}=1$ is used for estimation at each point in time. \\ \noindent
The indicators are used to select the correct period in which the individual-specific moment condition shall be used for estimation. Since $\tilde{N}_{it}=0$ in any period except the one  to be employed, hence I may simply multiply the individual- and time-specific moment conditions with the individual- and time-specific indicator variables and sum up over time to obtain the desired individual-specific moment condition, which is
\[ g_i= \sum_{t=1}^T \tilde{N}_{it}
\big(Y_{it}-\mu_{it}(p,q,S_0,G) \big) 
\]
Since the term in brackets only becomes relevant if $\tilde{N}_{it}=1$, I need to derive the mean of the outcome, conditional on $\tilde{N}_{it}=1$.
\[  \sum_{t=1}^T \tilde{N}_{it} \mu_{it}=
 \sum_{t=1}^T \tilde{N}_{it} E[Y_{it}|p,q,S_0,G]=
 E[Y_{it}|\tilde{N}_{it}=1,p,q,S_0,G]
\]
\noindent
Since $Y_{it}$ is binary, thus \[E[Y_{it}|\tilde{N}_{it}=1,p,q,S_0,G]=P(Y_{it}=1|\tilde{N}_{it}=1,p,q,S_0,G) \] 
Observe that the individual must enter period $t-2$ as uninformed because period $t-1$ is the first period in which she can be reached by the information. Therefore, conditioning on $\tilde{N}_{it}=1$ incorporates the information that $S_{i(t-2)}=0$. 
\[ P(Y_{it}=1| \tilde{N}_{it}=1,p,q,S_0,G)=
P(Y_{it}=1|S_{i(t-2)}=0, \tilde{N}_{it}=1,p,q,S_0,G)
\]
The second piece of information (besides not being informed in period $t-2$) that is incorporated in $\tilde{N}_{it}=1$ is that the individual has a positive probability to receive the information in period $t-1$, but a zero probability to receive the information any earlier. Only if a switch in information status has occurred in period $t-1$, a switch in the outcome is possible in period $t$. Consequently, the probability of the outcome being one can only be established conditioning on the agent's information status in the last two periods. 
 Whether or not the agent received the information in the exchange of period $t-1$ is not observable, hence I employ again the Law of Total Probability to obtain
 \[  P(Y_{it}=1|S_{i(t-2)}=0, \tilde{N}_{it}=1,p,q,S_0,G)=\]
\[ \sum_{s_{i,(t-1)}=0,1} \Big\{
P(Y_{it}=1|S_{i(t-1)}=s_{i(t-1)},S_{i(t-2)}=0, \tilde{N}_{it}=1,p,q,S_0,G) \times
\]
\[ P(S_{i(t-1)}=s_{i(t-1)}|S_{i(t-2)}=0,\tilde{N}_{it}=1, p,q,S_0,G) \Big\} \]
Note that the outcome in period $t$ depends only on the parameter $p$ and on whether or not the individual has been newly informed in the previous period. As a consequence, it is sufficient to condition on $S_{i,(t-1)}$, which takes the value one if the information has been received and zero if not, $S_{i(t-2)}$, which is known to be zero, and the model parameter $p$.
\[ P(Y_{it}=1|S_{i(t-1)}=s_{i(t-1)},S_{i(t-2)}=0, \tilde{N}_{it}=1,p,q,S_0,G) = \]
\[ P(Y_{it}=1|S_{i(t-1)}=s_{i(t-1)},S_{i(t-2)}=0,p) \]
\[ \mbox{\it for  }
s_{i(t-1)}=[0;1]
\] and therefore 
\[P(Y_{it}=1| \tilde{N}_{it}=1,p,q,S_0,G)=\]
\[ P(Y_{it}=1|S_{i(t-1)}=0,S_{i(t-2)}=0,p)P(S_{i(t-1)}=0|S_{i(t-2)}=0,\tilde{N}_{it}=1,
p,q,S_0,G) + \] 
\[ P(Y_{it}=1|S_{i(t-1)}=1,S_{i(t-2)}=0,p)P(S_{i(t-1)}=1|S_{i(t-2)}=0,\tilde{N}_{it}=1,p,q,S_0,G)  \]
Observe that the probability to participate in period $t$ is zero if the individual did not receive the information, hence the first term cancels. Note further that conditional on being newly informed, the individual's probability to participate is independent of any other agent's information or participation status and fixed at $p$. Thus 
\[P(Y_{it}=1| \tilde{N}_{it}=1,p,q,S_0,G)=\]
\[  P(Y_{it}=1|S_{i(t-2)}=0,S_{i(t-1)}=1,p)P(S_{i(t-1)}=1|S_{i(t-2)}=0,\tilde{N}_{it}=1,p,q,S_0,G) \] 
\[=p P(S_{i(t-1)}=1 |S_{i(t-2)}=0,\tilde{N}_{it}=1,p,q,S_0) \]
This shows that the unconditional expected value of the agent's outcome in the first period in which she potentially faces a choice is the parameter $p$ multiplied by the agent's probability to receive the information at the first possible occasion and I now seek to derive a closed-form expression for the latter. The agent's probability to receive the information in any period $t$ potentially depends on the information status of all other agents in period $t-2$. This is because any agent $j$ who enters period $t-1$ as informed (e.g. $S_{j(t-2)}=1) $ and who is linked to $i$ may send her the information. \\
\noindent Define the individual's information reception probability at time $t-1$, conditional on the information status vector of all villagers in period $t-2$ as
\begin{equation}
    r_{i(t-1)}(q,S_{t-2})=
1- \prod_{j=1}^N ( 1-q g_{ij} S_{j(t-2)}) \label{rr}
\end{equation}
Observe that $j$ can send to $i$ only if the two agents are linked $g_{ij}=1$ and further $j$ enters the period informed $S_{j(t-2)}=1$. Given both of these conditions are fulfilled, then the probability that the information spreads over a particular link is by assumption $q$. As a consequence, one minus the product of all three terms results in the probability of $j$ {\it not} sending to $i$.
Intuitively, $i$ receives the information whenever somebody sends it, thus the counterfactual (nobody sends it, the product of all relevant agents not sending it) can be used to calculate $r_{it}$.  
\\
\noindent
Information reception probabilities of the IPs are one, as they are known to be informed. For all other agents
\[ E[r_{i(t-1)}(q)]=
\sum_{s_{t-2}} r_{i(t-1)}(q,s_{t-2})P(S_{t-2}=s_{t-2},p,q,S_0,G) \]
Where I again used the Law of Total Probability: with $S_t$ (the information status of all villagers at point in time $t$) being unobserved after period $t=0$, thus calculating information reception probabilities hence implies integrating out the information status variables of all individuals that are linked to the agent. This technique is regularly employed in latent variable models, what makes its application challenging for the model at hand is the multitude of possible realizations $s_{t-2}$ of the random vector $S_{t-2}$. 
Further, the information status of all villagers in period $t-2$ in turn depends on the information status of their neighbours in the preceding period, which again is a function of the precedent information status vector. This argumentation can be applied until one ultimately ends up with the IPs, who are known to be informed. 
Because in each period there are a multitude of possibilities by which the information could have reached an agent, it is necessary to integrate out the entire random matrix $S_{1:t-2}$, comprising the information statuses of all villagers in all preceding time periods. Hence
\[ E[r_{i(t-1)}(q)]=
\sum_{s_{1:t-2}} r_{i(t-1)}(q,s_{1:t-2})
P(S_{1:t-2}=s_{1:t-2},p,q,S_0,G) \]
an expression that would have to be derived by iterated conditioning and then applying yet again the law of total probability. With every agent having two information statuses, the number of information scenarios (i.e. realizations of $s_{1:t-1}$) grows exponentially in the number of agents. This demonstrates that - as claimed in the introduction - calculating information reception probabilities becomes prohibitively costly even for intermediate time horizons and village sizes. \\
\noindent The probability of the agent entering period $t$ as informed in principle depends on all information exchanges that have been taking place beforehand. However, conditional on $\tilde{N}_{it}=1$, only the last information exchange needs to be considered as it was agent $i's$ first opportunity to be informed. Thus 
\[ P(S_{i(t-1)}=1 |S_{i(t-2)}=0,\tilde{N}_{it}=1,p,q,S_0,G) = E[r_{i(t-1)}(q)] \]
Then 
\[ \mu_{it} (p,q,S_0,G|\tilde{N}_{it}=1)=p E[ r_{i(t-1)}(q)] \]
And hence the moment conditions employed are
\[
   g_i(p,q,S_0,G)= \sum_{t=1}^T
\tilde{N}_{it} ( Y^i_t-p E[r_{i(t-1)}(q)])
\]
\[ E[g_i(p,q,S_0,G) ]=0 \]
Note that the mean formula $\mu_{it}$ was established conditioning on $\tilde{N}_{it}=1$ and hence is correct in only that one precise period. 
Since the indicator vector $\tilde{N}_{it}$ will be one in only one specific time period (the first in which the individual can be reached by the information), summing over time periods correctly selects the period for which the formula is applicable. 
\\ \noindent Naturally, other moment conditions can be derived. However, for one-parameter distributions, all higher order moments are functions of the mean and as such, the above derivations remain relevant. \\ \noindent
Having said that means are individual and time period specific is however slightly sloppy: indeed they are time-period and link-portfolio specific, since $r$ is pinned down by $G$ and $S_0$. Therefore, there are potentially multiple individuals with the same mean. This will be important for consistency of any estimator that is established on the basis of the vector $g_1,...,g_N$: as $N$ goes to infinity, every possible link portfolio will be observed in the data sufficiently often for any function that is established by summing up functions of the $g_i's$ to converge to its expected value.
% The system is over-identified and hence aggregate the moment conditions using a simple weighting function. The optimal weighting function is derived in later sections. Note that it appears now that we have $N$ individuals and $N$ moments, since each individual contributes to the moment conditions just once. 
% Having said that means are individual and time period specific is however slightly sloppy: indeed there are time-period and link-portfolio specific, since $r$ is pinned down by $G$. Therefore, there are potentially multiple individuals with the same mean. Above I choose to compute and use the individual-specific moments. An alternative would be to first aggregate the moments for all agents with identical means and then to use this set of (less then N) moments. The two proceedings are asymptotically identical since, as shall be proven below, the Covariance over individuals that at the same time enter the information radius is zero. As a consequence, the second proceeding is computationally substantially more complicated without adding a clear benefit.    

\subsection{Information Reception Probabilities}

Evaluating the individual-specific moment conditions above still requires an expression for $\tilde{N}_{it} E[r_{i(t-1)}(q)]$. While \eqref{rr} gives a formula how to compute the reception rate $r_{i(t-1)}$ given a specific realisation of $S_{t-2}$, the puzzle of how to circumvent the cumbersome computations of integrating out the random matrix $S_{1:t-2}$ still remains to be solved. This is the aim of this subsection. \\
\noindent
For IP-neighbours, calculating the information reception rates is straightforward. With $S_0$ being known, I can simply apply \eqref{rr}. 
Let $g_{ij}=1$ if $i$ and $j$ are linked and zero otherwise, then
\[ \Bar{r}_{i1}(q)=1- \prod_{ \tilde{N}_{j1}=1
} (1-q g_{ij}) \]
The information status vector of all individuals jointly fulfills the Markov property: conditioning on the last period's information status vector, I can calculate the current period's information reception probabilities. However, in order to obtain the unconditional reception rates, the latent past information status vector would have to be integrated out using once again the law of total probability. The number of information scenarios to be considered and hence the number of summands increases exponentially in the number of agents and time periods, thus making the calculation prohibitively time intense. 
The issue can be resolved in period two and three: given the choice of moment conditions, I can replace the past latent variables by their expectation (which is the respective individual's reception rate). This leads to the correct reception rate exclusively in the first period in which an agent can be reached by the information. For the second and in some cases the third information exchange, the same formula can be used as for the direct IP-neighbours. 
The following Theorem, a proof of which can be found in the Appendix, summarizes these findings. 

\begin{theorem}
\label{N2}
For individuals that are two links away from the information injection, the rates obtained from
replacing the individuals' neighbours' latent variables with their expected values will, in period 2 (the first period in which an agent can be reached), coincide with the rates that would be obtained from fully integrating out the random latent variables. \\
\noindent For individuals that are three links away from the information injection, the rates obtained from
replacing the individuals' neighbours' latent variables with their expected values will, in period 3 (the first period in which an agent can be reached), coincide with the rates that would be obtained from fully integrating out the random latent variables if there are no circles involving only non-IP non-participants on the way from the IP to the individual.
\end{theorem}
\noindent 
I now apply Theorem 1 to the second information exchange. Precisely, the random variables ($S_{j1}$) are replaced by the reception rates ($\bar{r}_{j1})$ in \eqref{rr}. 

\[ \Bar{r}_{i2}(q)= 1-\prod_{ \tilde{N}_{k2}=1
} (1- \Bar{r}_{k1}(q) q g_{ik}) \]
\noindent
For the third information exchange (used for the moments in period four), it is also not necessary to fully integrate out the past information status matrix. Here, however, two different formulas are available, depending on the final agent's link portfolio. These formulas are applied in accordance with the following Theorem.  

\begin{theorem}
\label{N3}
For individuals that are three links away from the information injection and for whom there are circles involving only non-IP non-participants on the way from the IP to the individual, the rates obtained from replacing the individual's neighbours' latent variables with their expected values and using an adapted formula will, in period 3 (the first period in which an agent can be reached), coincide with the rates that would be obtained from fully integrating out the random latent variables if and only if their direct neighbours do not have several friends in common that are directly linked to an initially informed individual, i.e.
\end{theorem}

\[ \Bar{r}_{i3}(q)= 1-\prod_{ \tilde{N}_{l3}=1} (1- \Bar{r}_{l2}(q) q g_{il}) 
\hspace{.45cm}
\mbox{if Theorem 1 applies}
\]
\[ \Bar{r}_{i3}(q)= 1- \prod_{ \tilde{N}_{k2}=1 } ( 1- \Bar{r}_{1k}(q) 
(1- \prod_{ \tilde{N}_{l3}=1 } (1-q^2 g_{kl}g_{li})))
\hspace{.45cm}
\mbox{if Theorem 3 applies}
\]
\noindent 
The first formula again uses Theorem 1 and replaces the (scenario specific) information status of $i's$ neighbours by their information reception probabilities in \eqref{rr}. \\
\noindent The last formula merits some further explanation. Agent $K$, who is directly linked to an IP is two links away from agent $i$ and there may be multiple paths between them, each passing through a specific intermediate agent $l$ (such that $g_{kl}g_{li}=1$). Since these are lines, the possibility of the information reaching $i$ through any one of them is simply $q^2$. Consequently, the probability of the information {\it not} reaching $i$ through that path is $1-q^2$ and if we take this to the power of the number of paths, I obtain the probability of agent $i$ not being informed through any of them. This conveniently facilitates the computation of agent $i$ being reached through any (one or several) paths from agent $k$ to agent $i$ as one minus the probability of her being reached through none. Since no path can frequent an intermediate agent twice by the condition of Theorem 2, they can be treated as independent. Finally, for agent $i$ to be informed through a path originating at agent $k$, the latter must have received the information in the first period. \\
\noindent 
It is crucial to take notice that all these rates are non-random, not scenario specific, but deterministic functions, pinned down by the network and the information initiation. \\ \noindent
Recapitulate that $r_{i(t-1)}(S_{t-2},q)$ is individual $i's$ information reception probability in period $t-1$ given the previous information scenario $S_{t-2}$. The crucial insight that drastically facilitates computations and hence makes exact estimation possible is that
\[
E[r_{i(t-1)}(q)|\tilde{N}_{it}=1]= \sum_{t=1}^T
\tilde{N}_{it} E[r_{i(t-1)}(q)]= 
\sum_{t=1}^T
\tilde{N}_{it}
\Bar{r}_{i(t-1)}(q)\] 
where the expectation is taken by integrating out the information scenarios. This is indeed what is stated in Theorems 1 and 2. \\
\noindent The aforementioned formulas are correct for the first period in which the agent can be reached by the information. 
For period 1, this is trivial: the IPs are known and there is no randomness. For period 2 and 3, a proof is provided in the Appendix. 
The moment conditions become 
\begin{equation}
   g_i(p,q,S_0,G)= \sum_{t=1}^T
\tilde{N}_{it} ( Y_{it}-p \Bar{r}_{i(t-1)} )
\label{MC}
\end{equation}
Naturally, in any period in which $\tilde{N}_{it}=0$, replacing the expected value of the individual's information reception probability with the estimate computed on the basis of Theorem 1 or 2 is incorrect, but since $\tilde{N}_{it}=1$ only for the one period in which using the simple formulas leads to a correct calculation, hence this is irrelevant. 
Exact estimation is thus feasible if, at each point in time, only a sub-sample of individuals is chosen to contribute to the objective function, namely those for which the rates can be straightforwardly obtained correctly. This implies that each individual contributes to the objective function just once. 
%Given that at each point in time one  disposes over one moment condition per link pattern, the parameters are identified even with only two individuals that have distinct linkages to the information. 
Given that the reception rates used are correct, there is no estimation bias.

\subsection{Within-Village Correlation}

This subsection explains how correlation in outcomes arises within the village. The result of this correlation is that any moment-based estimator exhibits a co-variance that is hard to evaluate or estimate. When villages are numerous, this correlation has minor effects, yet given a small to moderate number of villages, it impedes a fully efficient estimation. \\
\noindent
Correlation arises through the information reception probabilities. 
For any two agents residing in different villages, the moments are per definition uncorrelated. Further, IP moments are uncorrelated within and across villages. \\ \noindent  
Assume that agent $i$ and agent $j$ are non-IP individuals residing in the same village. 
Below I will use some very simple examples to illustrate how correlation of individual moment conditions arise within the village. 
I derive the covariation in individual moment conditions using formula \eqref{MC}.
\\ \noindent
\[ E[g_i(p,q,G,S_0)g_j(p,q,G,S_0)]= \]
\[ E\Bigg[ 
\Big(
\sum_{t=1}^T (Y_{it} -p \bar{r}_{i(t-1)})\tilde{N}_{it} \Big) \Big(
\sum_{t=1}^T (Y_{jt} -p \bar{r}_{j(t-1)})\tilde{N}_{jt} \Big) \Bigg] = \]
\[ E\Bigg[ \Big(
\sum_{t=1}^T (Y_{it})\tilde{N}_{it} \Big) \Big(
\sum_{t=1}^T (Y_{jt})
\tilde{N}_{jt}\Big) \Bigg]  -
p^2 
\Big(\sum_{t=1}^T \bar{r}_{i(t-1)}\tilde{N}_{it} \Big) \Big( 
\sum_{t=1}^T \bar{r}_{j(t-1)}\tilde{N}_{jt} \Big)
\] 
Assume that $\tilde{N}_{it} = \tilde{N}_{jt} \forall t$ e.g. both agents' first opportunity to receive the information is in the same period. Assume that $\tilde{N}_{i3}=\tilde{N}_{j3}=1$, e.g. both are indirectly linked to the information and that each disposes over exactly one IP connection, which passes through one and the same intermediate agent $k$. 
This is the set-up depicted in village graph 1.

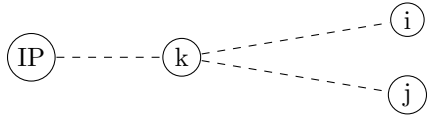
\begin{figure}[ht!]
    \centering
    \caption{Village Graph 1}    
\begin{tikzpicture}
\path
(0,0) node (1) [shape=circle,draw, inner sep=2,outer sep=.5] {IP }
(2,0) node (2) [shape=circle,draw, inner sep=2,outer sep=.5] {k }
(5,.5) node (4) [shape=circle,draw, inner sep=2,outer sep=.5] {i }
(5,-.5) node (5) [shape=circle,draw, inner sep=2,outer sep=.5] {j }
;
\draw [dashed] (1)--(2);
\draw [dashed] (2)--(4);
\draw [dashed] (2)--(5);
\end{tikzpicture}
\end{figure}

\noindent
Then 
\[ 
\sum_{t=1}^T \bar{r}_{i(t-1)} \tilde{N}_{it}=
\sum_{t=1}^T \bar{r}_{j(t-1)} \tilde{N}_{jt}=
\bar{r}_{k1} q
\]
Hence
\[ E[g_i(p,q,G,S_0)g_j(p,q,G,S_0)]= 
E[ Y_{i3} Y_{j3}] -p^2 (\bar{r}_{k1} q)^2
\]
Since $Y$ takes only the values 0 and 1 hence 
\[E[ Y_{i3} Y_{j3}]= P[Y_{i3}=1 \cap Y_{j3}=1] 
\] and since they only have one neighbour linked to the information (namely agent $k$) this is 
\[ 
P[Y_{i3}=1 \cap Y_{j3}=1]= \]
\[ P[Y_{i3}=1 \cap Y_{j3}=1| S_{k1}=1]P(S_{k1}=1)+
P[Y_{i3}=1 \cap Y_{j3}=1| S_{k1}=0]P(S_{k1}=0)
\]
\[ =
P[Y_{i3}=1 \cap Y_{j3}=1| S_{i2}=1 \cap S_{j2}=1 ] P(S_{i2}=1 \cap S_{j2}=1|S_{k1}=1)
P(S_{k1}=1)
\] The outcomes can only both be one if both agents are informed. \\ \noindent 
Conditional on their common friend $k$ being informed, the probability of $k$ sending to $i$ is independent of the probability of $k$ sending to $j$, hence 
\[ P(S_{i2}=1 \cap S_{j2}=1|S_{k1}=1)=P(S_{i2}=1|S_{k1}=1)P( S_{j2}=1|S_{k1}=1) \]
hence
\[ 
P[Y_{i3}=1 \cap Y_{j3}=1| S_{i2}=1 \cap S_{j2}=1 ] P(S_{i2}=1 |S_{k1}=1)
P(S_{i2}=1 |S_{k1}=1)
P(S_{k1}=1)
\]
\[ = p^2 q^2 \bar{r}^k_1 \neq p^2 (\bar{r}^k_1 q)^2
\] and thus 
\[ E[g_i(p,q,G,S_0)g_j(p,q,G,S_0)] \neq 0 \]
Correlation between moments of agents with the same set of indicators $\tilde{N}$ arises through friends in common. This is because 
\[ E[ r_{i(t-1)}(q) ]= 
 \sum_{s_{0:t-2}} r_{i(t-1)}(q,S_{0:t-2})
P(S_{1:t-2}|q,G,S_0)
\] hence friends in common show up in both agents' expected value, creating correlation in their outcomes.
If on the other hand $i$ and $j$ were linked to different intermediate agents (as depicted in village graph 2), then their outcomes would be independent. 

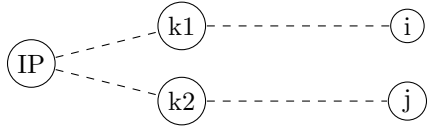
\begin{figure}[ht!]
    \centering
    \caption{Village Graph 2}    
\begin{tikzpicture}
\path
(0,0) node (1) [shape=circle,draw, inner sep=2,outer sep=.5] {IP }
(2,.5) node (2) [shape=circle,draw, inner sep=2,outer sep=.5] {k1 }
(2,-.5) node (3) [shape=circle,draw, inner sep=2,outer sep=.5] {k2}
(5,.5) node (4) [shape=circle,draw, inner sep=2,outer sep=.5] {i }
(5,-.5) node (5) [shape=circle,draw, inner sep=2,outer sep=.5] {j }
;
\draw [dashed] (1)--(2);
\draw [dashed] (1)--(3);
\draw [dashed] (2)--(4);
\draw [dashed] (3)--(5);
\end{tikzpicture}
\end{figure}

\noindent
Then 
\[  
E\Bigg[ 
\sum_{t=1}^T (Y_{it})\tilde{N}_{it} 
\sum_{t=1}^T (Y_{jt})\tilde{N}_{jt} \Bigg]
= \]
\[  
E\Bigg[ 
\sum_{t=1}^T (Y_{it})\tilde{N}_{it}  \Bigg] E \Bigg[
\sum_{t=1}^T (Y_{jt})\tilde{N}_{jt} \Bigg]
\]
\[ 
E\Bigg[ 
\sum_{t=1}^T (Y_{it})\tilde{N}_{it} \Bigg] =
p \bar{r}_{i2}= p \bar{r}_{k1,1} (q)
= \sum_{t=1}^T \bar{r}_{it}\tilde{N}_{it}
\]
\[
E\Bigg[ 
\sum_{t=1}^T (Y_{jt})\tilde{N}_{jt} \Bigg] = p \bar{r}_{k2,1}(q)
= \sum_{t=1}^T (\bar{r}_{jt})\tilde{N}_{jt}
\]
and thus 
\[ E[g_i(p,q,G,S_0)g_j(p,q,G,S_0)] 
=
0 \]
\\
\noindent Assume now that $ \exists t \hspace{.1cm}
s. th.\tilde{N}_{it} \neq \tilde{N}_{jt} $ such that one agent has the opportunity to receive the information earlier than the other. 
Again, I use a very simple case for illustration. 
Assume that $i$ is linked directly to an IP and that $j$ only disposes over one link to an IP, namely the one passing through $i$. Then 
\[ 
E\Bigg[ 
\sum_{t=1}^T (Y_{it})\tilde{N}_{it} \Bigg] = 
p \bar{r}_{i1} 
\] and 
\[ 
E\Bigg[ 
\sum_{t=1}^T (Y_{jt})\tilde{N}_{jt} \Bigg] = 
p \bar{r}_{i1} q 
\]
hence 
\[ E[g_i(p,q,G,S_0)g_j(p,q,G,S_0)]= 
E[ Y_{i2} Y_{j3}] -p^2 (\bar{r}^i_1 )^2 q
\]
But \[
E[ Y_{i2} Y_{j3}] =
P[Y_{i2}=1 \cap Y_{j3}=1|S_{i1}=1 ]P(S_{i1}=1) \]
\[+ P[Y_{i2}=1 \cap Y_{j3}=1|S_{i1}=0 ]P(S_{i1}=0)
\]
\[= P[Y_{i2}=1 \cap Y_{j3}=1|S_{i1}=1 ]P(S_{i1}=1) \]
since $i$ must be informed for her to have a non-zero probability to participate and for her to pass on the information to $j$. 
Since conditional on being informed, participation decisions are independent hence 
\[ P[Y_{i2}=1 \cap Y_{j3}=1|S_{i1}=1 ]P(S_{i1}=1) =
\]
\[  P[Y_{i2}=1 |S_{i1}=1 ]
 P[Y_{j3}=1 |S_{j2}=1 ] P(S_{j2}=1|S_{i1}=1)
P(S_{i1}=1)= \]
\[ p^2 P(S_{j2}=1|S_{i1}=1)
P(S_{i1}=1) \]
but $i$ is $j's$ only source of information, so
\[P(S_{j2}=1|S_{i1}=1)=q\] hence
\[ E[ Y_{i2} Y_{j3}] = p^2 q \bar{r}_{i1} \neq p^2 (\bar{r}_{i1} )^2 q \]
Implying that 
\[ E[g_i(p,q,G,S_0)g_j(p,q,G,S_0)] \neq 0\] 
Correlation in this case arises due to $i$ and $j$ being connected. 
\\ \noindent
Summing up, correlation arises through the information reception probabilities hence the means of the outcomes. Since for IPs, reception probabilities are non-random, thus per definition IP moments are uncorrelated with any other agents' moment, be it in the same or in a different village.

\subsection{Parameter Identification from Individual Moment Conditions}

As demonstrated above, the individual moment conditions are linear in $p$, but nonlinear in $q$. A basic precondition for identification is that there is some variation in the link portfolios, hence the information reception rates. Without such a variation, only the product of $p r(q)$ was identified. \\ \noindent
From the model on the other hand, it is always guaranteed that the sample includes at least one IP. For this agent, the information reception rate is one (she is sure to be informed). As a consequence
\[ E[Y_{i1}|\tilde{N}_{i1}=1 ]= p_0 \]
and hence the set of IPs can be used to identify $p$. 
With
\[ E \Bigg[ \sum_{t=1}^T (Y_{it})|\tilde{N}_{i1}=0 \Bigg]=
E \Bigg[ \sum_{t=2}^T (Y_{it})\tilde{N}_{it} \Bigg]
= p_0 \sum_{t=1}^T  E[r_{i(t-1)}(q)] \tilde{N}_{it} =\]
\[ p_0 \sum_{t=1}^T 
\bar{r}_{i(t-1)}(q_0) \tilde{N}_{it}
\]
obviously, the existence of non-IPs is necessary for the identification of $q$. When all non-IPs are either participants or non-participants, $q$ will not be identified in the interior of the parameter space and instead, since $0<q<1$, the estimate will be set to respectively the minimal or maximal value allowed. Disposing over at least one non-IP participant and one non-IP non-participant are thus a necessary condition for identifying $q$ (in the sense of not obtaining a corner solution).  
\\ \noindent
In the following, I propose two estimators: one that does not aggregate individual moment conditions at all and one that aggregates only IPs and non-IPs. 
\\
\noindent
The non-aggregated objective function is 
\[ 
\hat{Q}_{na}= N^{-1} \sum_{i=1}^N g_i^2
\]
while the two-moment objective function is 
\[ 
\hat{Q}_{2m}= (N_1^{-1} \sum_{i=1}^N g_i \tilde{N}_{1i})^2+
(N_2^{-1} \sum_{i=1}^N g_i (1-\tilde{N}_{1i}))^2
\]
With $N_1$ and $N_2$ denoting the number of IPs and non-IPs, respectively. 
The two-moment objective function has the advantage that the information on the sign of the outcome's deviation from its unconditional mean is preserved, but the inconvenience that within-village correlation is amplified, since when agents deviate in the same direction, the square of the sum of deviations exceeds the sum of the squares of deviations. \\
\noindent
The relationship between the two objective functions can be expressed as 
\[ \hat{Q}_{2m} \approx
\frac{1}{N}
\hat{Q}_{na} + 
\frac{2}{N_2^2} \sum_{i=1}^N \sum_{j=1}^N g_i g_j \] 
If the co-variation in individual-specific moments converges to zero as $N$ goes to infinity (which is what one expects), the two objective functions produce the same first-order-conditions, hence estimates. 
Since the latter term converges to zero when more villages are added, the estimates resulting from both functions are asymptotically equivalent but may exhibit different finite sample properties.

\section{The Non-aggregated Estimator}

\subsection{The Objective Function}

If aggregation within the village potentially results in small-sample bias and aggregation across villages is computationally costly, an idea would be to simply not aggregate at all. In the following, I call this the non-aggregated estimator. 
The sample moments are then 
\[ \hat{g}_i=g_i \]
Viewed as a GMM estimator, this would imply that we only have one observation per moment condition.
Stacking the individual moment conditions into a vector $g_s$ and individual moment conditions in village $v$ into a vector $g_v$, I obtain
 \[ \hat{Q}_{na}
= N^{-1} \hat{g}_s' \mathbf{I}_N \hat{g}_s 
\]
This is not a Method of Moments estimator in the sense it is generally defined in econometrics as I do not use sample moments but individual moments and I minimize the squared deviation of the observed variables from their unconditional expected values. 
This corresponds to a GMM estimator that uses agents of the same type (as defined by their link portfolio) to compose a type-specific moment condition. 
Since
\[\hat{Q}_{na}= N^{-1} \sum_{v=1}^V Q_{v,na}=
N^{-1}
\sum_{v=1}^V   g_v'\mathbf{I}_{N_v} g_v \]
hence this objective function has the substantial advantage that villages can be computed in parallel.
\\ \noindent In the following sections, I check all requirements for parameter identification and consistency. 

\subsubsection{The Limiting Function}

 Plugging in the expression for the individual-specific moments from above leads to
\[ \hat{Q}_{na}= N^{-1}
\sum_{i=1}^N \Big( 
\sum_{t=1}^T (Y_{it}-p \bar{r}_{i(t-1)}(q))\tilde{N}_{it} \Big)^2
\]
\[ 
Q_{0,na}=E[\hat{Q}_{na} ]=  E[g_i^2] =
E \Big[
\Big( \sum_{t=1}^T (Y_{it}-p \bar{r}_{i(t-1)}(q))\tilde{N}_{it} \Big)^2 \Big]
\]
It appears surprising at first sight that the sample equivalent of $Q_{0,na}$ should converge if each individual moment condition has a sample size of one. 
However, the objective function is established as a sum of the squared deviations of individual outcomes from their unconditional expected value. The latter is a function of the individual's link portfolio i.e. the lengths, but also concrete configuration of her link(s) to the IPs. There is a limited albeit large number of possible link portfolios to be encountered in the population. With several agents exhibiting the same link portfolio (thus unconditional mean), the objective function can be split into partial sums by aggregating agents of the same type.  
As the sample size increases, each link portfolio occurs sufficiently often for each partial sum to converge to its expected value and the fraction of agents of each type encountered in the sample converges to the respective proportion in the underlying population.  
Then $Q_{0,na}=E(Q_{na})$ can be seen as the squared difference between the outcome and its unconditional expected value in the first period in which an agent faces a choice for an agent who is representative for the underlying population. 
 Individuals can be segregated into $M$ types according to their link portfolio such that $I_{im}=1$ if agent $i$ has link portfolio $m$. 
Consequently
\[Q_{0,na}= E[g_i^2] = \sum_{m=1}^M E[g_i^2|I_{im}=1]P(I_{im}=1)\]

\noindent
Let $\mu_m$ be the unconditional expected value of the outcome for an agent with link portfolio $m$ in the first period in which she faces a choice and let $tm$ denote this particular period, (e.g. $\mu_m(p,q)=p \bar{r}_{m(tm-1)} $, then 
\[ 
E \Big[
\Big( \sum_{t=1}^T (Y_{it}-p \bar{r}_{i( t-1)}(q))\tilde{N}_{it} \Big)^2 |I_{im}=1 \Big]= \]
\[
E\Big[ \Big(Y_{i(tm)}- \mu_m \Big)^2|I_{im}=1\Big]=
E\Big[\Big(Y_{i(tm)}-p \bar{r}_{m(tm-1)}(q)\Big)^2|I_{im}=1\Big]
\]
I can rewrite 
\[ 
Q_{0,na}= \sum_{m=1}^M
E\Big[ \Big( Y_{i(tm)} - p \bar{r}_{m(tm-1)}(q) \Big)^2  |I_{im}=1\Big]
P(I_{im}=1)=
\]
\[  
\sum_{m=1}^M
\Bigg(
E\big[Y_{i(tm)}^2|I_{im}=1 \big]
 - 2 E\big[Y_{i(tm)}|I_{im}=1 \big]p \bar{r}_{m(tm-1)}(q) \]  
\[ + \big(p \bar{r}_{m(tm-1)}(q)\big)^2 \Big) 
P(I_{im}=1)
\]
Since $Y_{it}$ is binary, hence $E[Y_{it}^2]=E[Y_{it}]$. Further the expected value of the outcome given that an agent is of type $m$ is the type specific mean at the true parameters that generated the data, so 
\[ E[Y_{i(tm)}|I_{im}=1]= \mu_m (p_0,q_0)=
p_0 r_{m(tm-1)}(q_0) =
p_0 \bar{r}_{m(tm-1)}(q_0) 
\] and consequently
\[
Q_{0,na}= \sum_{m=1}^M
\Big(
p_0 \bar{r}_{m(tm-1)}(q_0)
-2 p_0 \bar{r}_{m(tm-1)}(q_0)p \bar{r}_{m(tm-1)}(q) \]
\[ +(p \bar{r}_{m(tm-1)}(q))^2 \Big)P(I_{im}=1)
\]
This also shows that the limiting function is continuous in the parameters.  

\subsection{Identification}

In order to demonstrate that $p_0,q_0$ maximises $Q_{0,na}$ I take the first order condition with respect to both parameters, which results in  
\[ Q_{0,na}= E[g_i^2] = \sum_{m=1}^M E[g_i^2|I_{im}=1]P(I_{im}=1)
\] \[
FOC_p: \sum_{m=1}^M
E[g_i|I_{im}=1]
\frac{\partial E[g_i|I_{im}=1]}{\partial p}
P(I_{im}=1)=0
\] \[ 
FOC_q: \sum_{m=1}^M
E[g_i|I_{im}=1]
\frac{\partial E[g_i|I_{im}=1]}{\partial q}
P(I_{im}=1)=0
\]
But since it has been derived above that 
\[ 
E[Y_{i(tm)}|I_{im}=1]= 
p_0 r_{m(tm-1)}(q_0)
\]
thus 
\[E[g_i(p_0,q_0)|I_{mi}=1]=
p_0 \bar{r}_{m(tm-1)}(q_0)
-p \bar{r}_{m(tm-1)}(q)=0
\]
hence $p_0,q_0$ is a solution to the FOCs. 

\subsection{Consistency}

What remains to be shown is that $\hat{Q}_{na}$ converges uniformly in probability to $Q_{0,na}$
Recall that 
\[ \hat{Q}_{na}= N^{-1}
\sum_{i=1}^N \Big( 
\sum_{t=1}^T (Y_{it}-p \bar{r}_{i(t-1)}(q))\tilde{N}_{it} \Big)^2
\]

\[ 
\hat{Q}_{na}= N^{-1}
\sum_{i=1}^N g_{i}^2
\]
If I sort the agents in my observed sample into the $M$ link portfolio-types from the above and use $N_m$ to denote the number of agents with link portfolio $m$ that are observed in the sample, this results in

\[
\hat{Q}_{na}= N^{-1} \sum_{i=1}^N
\sum_{m=1}^M 
\big( g_i I_{mi})^2
=
 N^{-1} 
\sum_{m=1}^M \sum_{i=1}^N
\big( g_i I_{mi})^2
\]
\[ =
 N^{-1} 
\sum_{m=1}^M 
\frac{N_m}{N_m}
\sum_{i=1}^N
 (g_i I_{im})^2=
 \frac{N_m}{N}
\sum_{i=1}^N
 \frac{1}{N_m}
 (g_i I_{im})^2=
\]
Then by a Law of Large numbers 
\[  
\sum_{i=1}^N
 \frac{1}{N_m}
 (g_i I_{im})^2 
\overset{p}{\to}
E[g_i^2|I_{im}=1]
\]
and 
\[ \frac{N_m}{N}
\overset{p}{\to} P(I_{im}=1)
\]
This is because the fraction of observed type-m agents converges to the probability that an individual that is randomly drawn from the underlying population is of type $m$. Further, since asymptotically every type is observed infinitesimally often, hence the average of the squared  deviations of type $m$ agents from their common mean converges to its type-specific expected value. As a consequence I can conclude that 
$\hat{Q}_{na}$ converges uniformly to $Q_{0,na}$. \\
\noindent Note that convergence occurs even though individual outcomes are correlated in the village. As long as variances and covariances are bounded, the LLN remains applicable. The boundedness of the covariances is assured by the assumption that an increase n the sample size is achieved by adding more villages, guaranteeing that the vast majority of pair of agents are uncorrelated as they reside in different villages.  

\subsection{Asymptotic Variance}

Let $\theta=(p,q)$ be the vector of model parameters. 
Let $\hat{S}(\theta)$ denote the column vector of first derivatives of the observed objective function with respect to $p$ and $q$ and $\hat{H}(\theta)$ the matrix of second order derivatives, then 
a mean-value expansion of the vector $\hat{S}$ leads to 
\[  
\hat{S}(\hat{\theta})=
\hat{S}(\theta_0)+ \hat{H}(\bar{\theta})(\hat{\theta}-\theta_0)=0
\]
and thus 
\[ 
(\hat{\theta}-\theta_0)=
\hat{H}(\bar{\theta})^{-1}\hat{S}(\theta_0)
\]
such that the asymptotic variance is
\[ V_{na}=
E[ 
\hat{H}( \theta_0)^{-1}\hat{S}(\theta_0)
\hat{S}(\theta_0)'
\hat{H}(\theta_0)^{-1}
] \]
Denote the matrix of partial derivatives of the moments conditions as
\[ D= \frac{\partial g_s}{ \partial \theta }
= \Big(
 \frac{\partial g_1}{ \partial \theta },...,
  \frac{\partial g_N}{ \partial \theta } \Big)
\] 
Observe that 
\[
E[\hat{S}(\theta) \hat{S} (\theta)']=
E \Bigg[
\Big(
\sum_{i=1}^N 
2  \frac{\partial g_i}{\partial \theta} g_i\Big)
\Big(
\sum_{i=1}^N 
2  \frac{\partial g_i}{\partial \theta} g_i\Big)'
\Bigg] = \]
\[
4
\begin{bmatrix}
E \Big[
\Big(
\sum_{i=1}^N 
 \frac{\partial g_i}{\partial p} g_i\Big)^2 \Big] &
E \Big[
\Big(
\sum_{i=1}^N 
  \frac{\partial g_i}{\partial p} g_i\Big)
\Big(
\sum_{i=1}^N 
  \frac{\partial g_i}{\partial q} g_i\Big) 
\Big] \\
E \Big[
\Big(
\sum_{i=1}^N 
 \frac{\partial g_i}{\partial p} g_i\Big) 
\Big(
\sum_{i=1}^N 
  \frac{\partial g_i}{\partial q} g_i\Big)
\Big]  & 
E \Big[
\Big(
\sum_{i=1}^N 
2  \frac{\partial g_i}{\partial q} g_i\Big)^2 \Big] 
\end{bmatrix}
\] 
\[ =E[ \hat{D}\hat{\Omega} \hat{D} ] \]
and 
\[
E[\hat{H}(\theta)]= \]
\[2
\begin{bmatrix}
E \Big[
\sum_{i=1}^N 
 \Big(\frac{\partial g_i}{\partial p} \Big)^2 
 + \sum_{i=1}^N \Big(
 \frac{\partial^2 g_i}{\partial^2 p} g_i \Big)
 \Big] & 
E \Big[
\sum_{i=1}^N 
 \frac{\partial g_i}{\partial p} 
 \frac{\partial g_i}{\partial q} 
 +\sum_{i=1}^N 
 \Big(  \frac{\partial^2 g_i}{\partial p \partial q} g_i \Big)
 \Big] \\
 E \Big[
\sum_{i=1}^N 
  \frac{\partial g_i}{\partial p} 
 \frac{\partial g_i}{\partial q} 
 + \Big( \sum_{i=1}^N 
 \frac{\partial^2 g_i}{\partial p \partial q} g_i \Big)
 \Big] &
 E \Big[
\sum_{i=1}^N 
 \Big(  \frac{\partial g_i}{\partial q} \Big)^2 
 + \sum_{i=1}^N 
 \Big(  \frac{\partial^2 g_i}{\partial^2 q} g_i \Big)
 \Big]
\end{bmatrix}
\]

\[
E[\hat{H}(\theta)]= 
2
\begin{bmatrix}
E \Big[
\sum_{i=1}^N 
 \Big(\frac{\partial g_i}{\partial p} \Big)^2 
  & 
E \Big[
\sum_{i=1}^N 
 \frac{\partial g_i}{\partial p} 
 \frac{\partial g_i}{\partial q} \Big]
  \\
 E \Big[
\sum_{i=1}^N 
  \frac{\partial g_i}{\partial p} 
 \frac{\partial g_i}{\partial q} \Big]
 &
 E \Big[
\sum_{i=1}^N 
 \Big(  \frac{\partial g_i}{\partial q} \Big)^2 \Big] 
\end{bmatrix}
\]
since $E[g_i]=0$.
Hence
\[ E(H(\theta))=
E[D'\mathbf{I}_N D]
\]
then the results from the above show that $V$ can equivalently be represented as 
\[ V=
(D'\mathbf{I}_N D)^{-1} D'\mathbf{I}_N \Omega  
\mathbf{I}_N D'
(D'\mathbf{I}_N D)^{-1}  
\] which again underlines that the non-aggregated estimator can be perceived as a GMM estimator. Note that since each moment condition has the same weight, the weights just cancel out. 
\\ \noindent
Because moments are correlated within a village, but not across villages, hence $\Omega$ is block diagonal. One block comprises the village specific covariance matrix $\Omega_v$. The latter has the individual variances on the diagonal and covariances between individual moments on the off diagonal elements.
\noindent
To construct a fully efficient estimator, all the off-diagonal elements would have to be specified or estimated. If villages were identical, $\Omega_v$ could be estimated and a fully efficient GMM estimate could be constructed. In principle, correlations are specific to any particular pair of agents and additional assumption would be needed to apply methods that facilitate the calculation of covariances in the presence of clusters.  
The alternative is to derive them in terms of the parameters. In the model at hand, this is possible: with $G$ being observed, all the correlations and hence all non-zero elements of $\Omega$ can be expressed in terms of $p$ and $q$. However, this is computationally prohibitively expensive: for any two agents in the same information radius, the number of friends in common as well as the friends' reception rates would have to be found while for any two individuals at different IP-distance, it would have to be verified whether one is on the path between the other and some IP. There would be $(N_v-IP)(N_v-IP-1)$ of such checks per village and each would be computationally cumbersome. As a consequence, specifying $\Omega$ correctly in terms of $p$ and $q$ is computationally unrealistic. 
 \\ \noindent
 Observe however that the sample variance is 
\[ \hat{V}=
(N^{-1}D'\mathbf{I}_N D)^{-1} N^{-2}D'\mathbf{I}_N \Omega  
\mathbf{I}_N D'
(N^{-1}D'\mathbf{I}_N D)^{-1}  
\] The matrix
\[ D'\mathbf{I}_N D 
\] does not involve covariances and an element of 
\[ ( D'\mathbf{I}_N \Omega  
\mathbf{I}_N D')_{11}=
\sum_{i=1}^N \big( \sum_{j=1}^N 
\frac{\partial g_j}{\partial p} \rho_{ij} \big) 
\frac{\partial g_i}{\partial p}
\]
\[ ( D'\mathbf{I}_N \Omega  
\mathbf{I}_N D')_{12}=D'\mathbf{I}_N \Omega  
\mathbf{I}_N D')_{21}=
\sum_{i=1}^N \left( \sum_{j=1}^N 
\frac{\partial g_j}{\partial p} \rho_{ij} \right) 
\frac{\partial g_i}{\partial q}
\]
\[ ( D'\mathbf{I}_N \Omega  
\mathbf{I}_N D')_{22}=
\sum_{i=1}^N \left( \sum_{j=1}^N 
\frac{\partial g_j}{\partial q} \rho_{ij} \right) 
\frac{\partial g_i}{\partial q}
\]
with $\rho_{ij}=E[g_i g_j] $ being the variance or co-variance respectively. Correlations within the villages are always positive. The inner sum in the expression above has at most $N_v$ non-zero terms. 
Assume that $N \rightarrow \infty$ by means of $V \rightarrow \infty$, e.g. we obtain more (not larger) villages. Observe that I can rewrite 
\[( D'\mathbf{I}_N \Omega  
\mathbf{I}_N D')_{11} 
= 
\sum_{v=1}^V 
\sum_{i=1}^{N_v} \left( \sum_{j=1}^{N_v}
\frac{\partial g_j}{\partial p} \rho_{ij} \right) 
\frac{\partial g_i}{\partial p}
\] and thus 
\[  N^{-2}
( D'\mathbf{I}_N \Omega  
\mathbf{I}_N D')_{11} 
= 
\sum_{v=1}^V 
\left(\frac{N_v}{N}\right)^2
\left(\frac{1}{N_v}\right)^2
\sum_{i=1}^{N_v} \left( \sum_{j=1}^{N_v}
\frac{\partial g_j}{\partial p} \rho_{ij} \right) 
\frac{\partial g_i}{\partial p} \]
Since there are at most $N_v$ positive correlations in village $v$ hence 
\[ 
\left(\frac{1}{N_v}\right)^2
\sum_{i=1}^{N_v} \left( \sum_{j=1}^{N_v}
\frac{\partial g_j}{\partial p} \rho_{ij} \right) 
\frac{\partial g_i}{\partial p} 
\] converges to its expected value. As such, with village sizes being fixed, the variance of the estimator does converge to zero, albeit at a rate that is slower than $N^{-1}$. To ease the exposition, assume villages have equal size then 
\[( D'\mathbf{I}_N \Omega  
\mathbf{I}_N D')_{11} 
= V \left(\frac{N_v}{N}\right)^2
\left(\frac{1}{N_v}\right)^2
\sum_{i=1}^{N_v} \left( \sum_{j=1}^{N_v}
\frac{\partial g_j}{\partial p} \rho_{ij} \right) 
\frac{\partial g_i}{\partial p} \] 
As $N \rightarrow \infty$ with $N_v$ being fixed and $V$ increasing less than proportional to $N$, thus 
\[ \hat{V} \rightarrow 0 \] albeit slower than $N^{-1}$. 
With fixed-$N_v$ asymptotics,  e.g.\  an increase in the sample size is achieved by adding more villages, the correlation in individual moments play a minor role. This is also be the case if village sizes are increased while holding the number of IPs constant and spreading them out. 
\\ \noindent 
Having said that there are at most $N_v^2$ elements in the village-specific sums (e.g.\ the correlation of each villagers' moment condition with herself and with all other villagers) does not mean that there are indeed this many positive elements as villagers may well be uncorrelated if they are not linked to the same IP and do not have friends in common. Presuming that correlations among moments play a negligible role, one could presume that $\Omega$ is approximately diagonal with individual-specific variances as diagonal elements. The presumed diagonal matrix $\Omega$ can be used as a weighting matrix in the moment estimation. 
Importantly, $E[\Omega]$ fails to be diagonal if there is at least one pair of non-IP agents residing in the same village with either one being on the path from the information to the other or both being linked to the same direct source of information. However, as the sample size increases, the error that is made by replacing  with a diagonal matrix becomes negligible and correlations within the village play a minor role.

\section{Two-moment GMM Estimator}

\subsection{The Objective Function}

Since it potentially leads to bias to aggregate the individuals of one village into one moment, a simple remedy is to aggregate all individuals into two sample moments, one comprising the IPs and the other all remaining agents.
The moments for IPs are uncorrelated with the moments of any other individual, even within the village. This is because correlation arises solely through information reception probabilities. IPs already have the information and by assumption conditional on being informed participation is an individual choice. Consequently, the IP-moment is thus composed of i.i.d.\ data. For the non-IP moment, individual moment conditions are positively correlated within the village as derived above. One may aggregate non IPs in an attempt to reduce small-sample bias, hoping that (correlated) positive deviations from the individual-specific means in one village may cancel with (correlated) negative deviations in another village. \\
\noindent
Since I decide to use only the IPs for the first and the non-IPs for the second moment, thus I choose the population moment conditions
\[ g_1= E[g_i |\tilde{N}_{1i}=1]\]
and 
\[ g_2=  E[g_i |\tilde{N}_{1i}=0]
\]
Each individual moment condition can be split into 
\begin{equation} \label{split}
   g_i= (Y_{i1}-p) \tilde{N}_{i1}+
\sum_{t=2}^T (Y_{it}-p \bar{r}_{i(t-1)}
) \tilde{N}_{it} 
\end{equation}
where I made use of the fact that whenever $\tilde{N}_{i1}=1$ (an individual faces a choice in period one, i.e.\ she is an IP), then $\bar{r}_{i0}=r_{i0}=1$ (she is known to be informed).
This shows that each individual contributes to one moment, 
i.e.\ the first moment is estimated from the sub-sample of IPs, the second from the sub-sample of non-IPs. 
Let $N_1$ denote the number of IPS and $N_2$ the number of non-IPs that are in the information radius. Then the sample moments are 
\[ 
\hat{g}_1= \frac{1}{N_1} \sum_{i=1}^N (Y_{1i}-p) \tilde{N}_{i1}
\]
\[
\hat{g}_2= \frac{1}{N_2} \sum_{i=1}^N 
\sum_{t=2}^T \big( 
Y_{it}- p \bar{r}_{i(t-1)} \big) \tilde{N}_{it}
\]
Note that summing up over all individuals is inconsequential as the index vector ($\tilde{N}_i$) conveniently guarantees that individual $i$ contributes only to one of the two moments and hence the sum in $\hat{g}_1$ (respectively $\hat{g}_2$) will always have $N_1$ (respectively $N_2$) terms. 
\[ 
\hat{Q}_a = (\hat{g}_1, \hat{g_2})' \mathbf{I}_2
(\hat{g}_1, \hat{g_2})
\]
Note that this is but one weighting choice. However, as shall be demonstrated below, as long as the first-order-condition (FOC) with respect to $q$ has a solution, the choice of the weighting matrix has no impact. As a robustness check, I computed the estimates using weights corresponding to the relative sample sizes, i.e.\ $w_1=\frac{N_1}{N}, w_2=\frac{N_2}{N}$. The estimates were identical.

\subsection{The limiting Function}

As mentioned above, the first moment consists of the moment conditions for the IPs, i.e.\ those individuals for which $\tilde{N}_{1i}=1$. 
\[ E(g_1)=  E[g_i |\tilde{N}_{i1}=1]=
E[(Y_{i1}-p) \tilde{N}_{i1}|\tilde{N}_{i1}=1)]
\] 
because for IPs, the second term in \eqref{split} is zero. 
\[ E(g_1)=  E[Y_{i1}|\tilde{N}_{i1}=1]-p
\] 
By the assumptions stated above, these individuals are known to be informed (i.e.\ $r_{i0}=1$), hence the expected value of their outcome is simply $p_0$ (i.e.\ the true probability that an informed individual participates).   
Thus 
\[ E[g_i|\tilde{N}_{i1}=1]=P(Y_{i1}=1|S_{i0}=1)-p=
p_0-p\] 
For all other individuals, the true unconditional expected value of the outcome consists of the probability to participate multiplied by the information reception probability. Since $\tilde{N}_{it}=1$ only in the period that directly follows the first information exchange in which $i$ could have been informed, hence 
\[ E(g_2)=
E[g_i|\tilde{N}_{i1}=0]=
E[
\sum_{t=2}^T (Y_{it}-p \bar{r}_{i(t-1)}(q))\tilde{N}_{it}
]\]
because for non-IPs, the first part in \eqref{split} is zero. By the same argument as above, I can segregate agents according to their link portfolio, i.e. information reception probabilities. For simplicity, denote type $m=1$ as the IPs, then 
\[
E[g_i|\tilde{N}_{i1}=0]=
\sum_{m=2}^M E[g_i|I_{im}=1]P(I_{im}=1|\tilde{N}_{i1}=0)
\] 
i.e.\ the expected value of the individual moment condition of a non-IP is calculated by the law of total probability taking into account how likely it is that a randomly drawn non-IP exhibits link portfolio $m$ (which is $P(I_{im}=1|\tilde{N}_{i1}=0)$). 
\\ \noindent 
Recalling that $tm$ is defined as the first period in which a type $m$ non-IP agent faces a choice then 
\[ 
E[g_{i}|I_{mi}=1]=p_0 \bar{r}_{m,tm-1}(q_0)-
p \bar{r}_{m,tm-1}(q)
\]
Therefore 
\[ E(g_2)=
\sum_{m=1}^M 
(p_0 \bar{r}_{m,tm}(q_0)-
p \bar{r}_{m,tm}(q))
P(I_{im}=1|\tilde{N}_{i1}=0)
\] 
\[ E[Q_a]=Q_{0,a}= E(g_1)^2+E(g_2)^2
\] 

\subsection{Identification}

Again it is necessary to show that the First-order Conditions (FOCs) of the limiting function are zero at the true parameter value. 
\[ 
FOC_p:
-2 E(g_1)
\frac{\partial E(g_1)}{\partial p}
- 2E(g_2)
\frac{\partial E(g_2)}{\partial p}=0
\Rightarrow
\] 
\[ 
 \big( p_0-p \big)+
\left(\sum_{m=1}^M 
(p_0 \bar{r}_{m,tm}(q_0)-
p \bar{r}_{m,tm}(q))
P(I_{im}=1|\tilde{N}_{i1}=0)\right) \]
\[\left(\sum_{m=1}^M \bar{r}_{m,tm}(q)
P(I_{im}=1|\tilde{N}_{i1}=0 )\right)
=0
\] 
\[
FOC_q:
-2
E(g_2)
\frac{\partial E(g_2)}{\partial q}= 0
\Rrightarrow
\] 
\[ 
\Bigg(\sum_{m=1}^M 
(p_0 \bar{r}_{m,tm}(q_0)-
p \bar{r}_{m,tm}(q))
P(I_{im}=1|\tilde{N}_{i1})=0) \Bigg)
\]
\[\left(\sum_{m=1}^M 
p \frac{\partial \bar{r}_{m,tm}(q)}{ \partial q}P(I_{im}=1|\tilde{N}_{i1}=0 ) \right)
=0
\]
Thus apparently, $p=p_0, q=q_0$ is a solution to the FOCs.
Note firstly that the estimate of $q$ is pinned down solely by means of the second moment. Since the FOC of $q$ can be expressed as 
\[FOC_q:
- 2 E(g_1) E \frac{\partial g_1}{ \partial q}
=0
\] 
Since the partial derivative is non-zero, if there is an exact solution to the FOCs of $q$, then the latter implies that the $pq$ combination is chosen such that $E(g_1)=0$. This pins down $p$ as a function of $q$ such that this condition is fulfilled. Substituting this into the FOCs for $p$ implies that in the latter, the second term is zero and as such, the first moment identifies $p$. This also highlights that is case of uniqueness, any weights chosen are irrelevant to the resulting estimate, which was confirmed in the Monte Carlo study. 

\subsection{Consistency}

Recalling that 
\[
\hat{g}_1= \frac{1}{N_1} \sum_{i=1}^N (Y_{1i}-p) \tilde{N}_{1i}
\]
As $N \rightarrow \infty$, also $N_1 \rightarrow \infty$, albeit at a smaller rate. I can express $N_1=N \frac{N_1}{N}=N w_1$ where $w_1$ is the fraction of $IPs$ in the underlying population. This implies that $\hat{g}_1$ converges to $E(g_1)$ by a Law of Large Numbers (LLN). \\
\noindent To see this note that 
\[
\hat{g}_1= \frac{N}{N_1}
\frac{1}{N}
\sum_{i=1}^N (Y_{1i}-p) \tilde{N}_{1i}\]
Then by a LLN
\[ 
\frac{1}{N} \sum_{i=1}^N (Y_{1i} -p)\tilde{N}_{1i}
\overset{p}{\to} E[Y_{i1} \tilde{N}_{i1}]
\]
By the Law of iterated expectation 
\[ E[(Y_{i1}-p) \tilde{N}_{i1}] 
= 
E[ E[(Y_{i1}-p) \tilde{N}_{i1}] |\tilde{N}_{i1}]=
E[ \tilde{N}_{i1}  E[Y_{i1}-p |\tilde{N}_{i1}]]
\]
Since $\tilde{N}_{i1}$ is binary hence 
\[
E[ \tilde{N}_{i1} E[Y_{i1}-p |\tilde{N}_{i1}]]=
 1 \times E[Y_{i1}-p |\tilde{N}_{i1}=1]P(\tilde{N}_{it}=1)+
 0 \times E[Y_{i1}-p |\tilde{N}_{i1}=0]P(\tilde{N}_{it}=0)\]
 \[ E[Y_{i1} -p|\tilde{N}_{i1}=1]P(\tilde{N}_{it}=1)
 = (p_0-p) P(\tilde{N}_{it}=1)
 \]
 Further by a LLN
 \[  \frac{N}{N_1} \overset{p}{\to} P(\tilde{N}_{it}=1)^{-1}\]
 This is because the reciprocal of the fraction of IPs encountered in the sample will converge to the probability that a randomly selected individual from the underlying population is an IP. As a consequence 
 \[ 
 \hat{g}_1 \overset{p}{\to} 
 P(\tilde{N}_{it}=1)^{-1} (p_0-p) P(\tilde{N}_{it}=1)=
 p_0-p=E(g_1)
 \]
\[
\hat{g}_2= \frac{1}{N_2} \sum_{i=1}^N 
\sum_{t=2}^T \big( 
Y_{it}- p \bar{r}_{i(t-1)} \big) \tilde{N}_{it}
\]
Which can be expressed as 
\[ 
\frac{1}{N_2} \sum_{m=2}^M 
(Y_{it}-p\bar{r}_{m,tm}(q)) I_{im}=
\frac{N}{N_2} \sum_{m=2}^M 
\frac{1}{N}
\sum_{i=1}^N 
(Y_{it}-p\bar{r}_{m,tm}(q)) I_{im}\]
Then we know that by a LLN 
\[ 
\frac{1}{N}
\sum_{i=1}^N 
(Y_{it}-p\bar{r}_{m,tm}(q)) I_{im}
 \overset{p}{\to} 
 E(Y_{it}-p \bar{r}_{m,tm-1}|I_{im}=1)P(I_{im}=1)=\]
\[ (p_0 \bar{r}_{m,tm-1}(q_0)-p \bar{r}_{m,tm-1}(q))P(I_{im}=1)
\]
and 
\[ \frac{N}{N_2} \overset{p}{\to} 
P(\tilde{N}_{i1}=0)^{-1}
\]
Finally since 
\[
P(I_{im}=1|\tilde{N}_{i1}=0)= 
\frac{P(I_{im}=1)}{P(\tilde{N}_{i1}=0)}
\]
Hence 
\[ 
\hat{g}_2 \overset{p}{\to} 
(p_0 \bar{r}_{m,tm-1}(q_0)-p \bar{r}_{m,tm-1}(q_0))P(I_{im}=1|\tilde{N}_{i1}=0)=
E(g_2)
\]
From which I can deduct that $\hat{Q}_a$ converges uniformly to $Q_{0,a}$. 
With the objective function converging to its expected value and the latter being minimized at $p=p_0,q=q_0$, we can thus conclude that the estimator is consistent. 

\subsection{Asymptotic Variance}

By the standard GMM results we have 
\[ V= (D'\mathbf{I}_2 D)^{-1} 
(D'\mathbf{I}_2 \Omega \mathbf{I}_2 D)
(D'\mathbf{I}_2 D)^{-1}
\]
Now $\Omega$ is diagonal, since IPs and non-IPs are uncorrelated. Consequently we have 
\[ \Omega_{11}=\frac{p_0(1-p_0)}{N_1} \]
\[ 
\Omega_{22}= E\Big[\Big( \frac{1}{N_2}
\sum_{i=1}^{N} g_i (1-\tilde{N}_{1i})\Big)
\Big( \frac{1}{N_2}
\sum_{i=1}^{N} g_i (1-\tilde{N}_{1i})\Big) \Big]
= \sum_{i=1}^{N_2} 
\sum_{j=1}^{N_2} 
\frac{\rho_{ij}}{N^2}
\]
With $\rho_{ij}$ being the Variance of $g_i$ if $i=j$ or the covariance between $i$ and $j$ if $i \neq j$. 
By the same argument as above, this can be re-written as 
\[ 
= 
\sum_{v=1}^V \frac{N_{2v}}{N_2}
\sum_{i=1}^{N_{2v}} 
\sum_{j=1}^{N_{2v}} 
\frac{\rho_{ij}}{N_v^2}
\]
which shows that the estimator also converges at a rate slower than $N^{-1}$. 

\section{Comparison}

From the first-order-condition of their respective limiting functions it is apparent that whenever the solution is unique, both estimation methods consistently identify it.  
In this case, the non-IPs identify $q$: $q$ is chosen such that the derivative of the objective function with respect to $q$  (which only depends on non-IPs) is set to zero. Since choosing $q$ such that the sum of deviations (hence the sum of individual non-IP moment conditions) is zero is a solution to the foc for $q$, therefore in case of uniqueness, it is the only solution in both cases. The IPs on the other hand then identify $p$: since $q=q(p)$ is chosen such that the moment conditions for non-IPs are zero for any $p$ satisfying the relationship that was derived from the foc for $q$, hence non-IPs do not contribute any further information relevant to the choice of $p$ beyond this relationship and it is the IPs that identify $p$. 

\subsection{Small Sample Properties}

Though both estimators are consistent given uniqueness, yet the finite sample estimates and also the estimates in case of multiplicity of solutions to the FOCs vary. Multiplicity does not appear to be an issue for the problem at hand, however, the difference in finite sample properties merits attention. Recalling that agents can be segregated into $M$ link portfolios, the first link portfolio denoting the IPs, then the finite sample first-order conditions are: \\
{\bf Non-aggregated estimator:}
\[  
FOC_p: \]
\[\frac{N_1}{N}\frac{1}{N_1}
\sum_{i=1}^N (Y_{i1}-p)\tilde{N}_{i1}+
\frac{N_2}{N}\frac{1}{N_2}
\sum_{i=1}^N
\sum_{t=2}^T
(Y_{it}- p \bar{r}_{i(t-1)})
\bar{r}_{i(t-1)}
\tilde{N}_{it}=0
\]
{\bf Two-moment estimator:}
\[  
FOC_p: \]
\[\frac{1}{N_1}
\sum_{i=1}^N (Y_{i1}-p)\tilde{N}_{i1}+
\Bigg(\frac{1}{N_2}
\sum_{i=1}^N
\sum_{t=2}^T
(Y_{it}- p \bar{r}_{i(t-1)})\tilde{N}_{it} \Bigg)
\Bigg( \frac{1}{N_2}
\sum_{i=1}^N
\sum_{t=2}^T
\bar{r}_{i(t-1)}
\tilde{N}_{it} \Bigg) =0
\]
{\bf Non-aggregated estimator:}
\[
FOC_q:\]
\[\frac{N_2}{N}\frac{1}{N_2}
\sum_{i=1}^N
\sum_{t=2}^T
(Y_{it}- p \bar{r}_{i(t-1)})
p \frac{\partial \bar{r}_{i(t-1)}}{\partial q}
\tilde{N}_{it}=0
\]
{\bf Two-moment estimator:}
\[
FOC_q:\]
\[ 
\Bigg(
\frac{1}{N_2}
\sum_{i=1}^N
\sum_{t=2}^T
(Y_{it}- p \bar{r}_{i(t-1)})
\tilde{N}_{it}\Bigg)
\Bigg(
\frac{1}{N_2}
\sum_{i=1}^N
\sum_{t=2}^T
p \frac{\partial \bar{r}_{i(t-1)}}{\partial q}
\tilde{N}_{it}\Bigg)=0
\]
Note that the multiplication by $\frac{N_1}{N}$ and $\frac{N_2}{N}$ would play a role only in case of non-uniqueness, which did not seem to be the case. \\ \noindent 
First, the difference concerns solely non-IPs. Second, for the FOC of $q$,  for the Two-Moment estimator, the term in the second bracket will naturally never be zero, hence the  Two-Moment-Estimator minimises the sum of deviations of all non-IPs from their unconditional expected value, giving equal weight to each agent (i.e. it minimises the term in the first brackets). The Non-aggregated estimator in contrast minimises a weighted sum of deviations, the weights being the partial derivative of the moment condition with respect to $q$. 
The latter is a function of the derivative of agents' information reception probabilities ($\frac{\partial r_{i(t-1)}}{\partial q}$). The function is decreasing and convex in the individual's degree hence implying that less weight is given to very well connected agents. 
For the foc of $p$, for the two-moment estimator, $q=q(p)$ has been chosen such that the entire non-IP term is as small as possible and it is thus predominantly the IP term that pins down $p$. For the non-aggregated estimator on the other hand, individual deviations from their expected value are weighted with the derivative of the moment condition with respect to $p$, which is the information reception rate. The latter is increasing in the individual degree with diminishing slope, implying that more weight is allocated to well-connected agents. \\
\noindent The FOCs thus show that the non-aggregated estimator takes into account also the marginal (as opposed to only the absolute) effect of a change in $p$ or $q$ on the individual moment conditions. 
\\ \noindent
Expressing 
\[ 
\hat{
\varepsilon}_i=
\sum_{t=1}^T(Y_{it}-p \bar{r}_{i,t-1})\tilde{N}_{it}
\]
as the actual deviation from the mean observed for individual $i$. It is apparent that 
\[ 
Q_{na}=
\frac{1}{N}
\sum_{i=1}^N
\hat{
\varepsilon}_i^2 
\] and 
\[ 
Q_{a}=(
\frac{1}{N_1}
\sum_{i=1}^N \hat{
\varepsilon}_i 
\tilde{N}_{i1})^2+
(\frac{1}{N_2}\sum_{i=1}^N \hat{
\varepsilon}_i 
(1-\tilde{N}_{i1}))^2 =
\]
\[ \frac{1}{N_1^2}
 \sum_{i=1}^N \hat{
\varepsilon}_i^2 \tilde{N}_{i1}
+ 
\frac{1}{N_1^2}
\sum_{i=1}^N 
\sum_{j \neq i}
\hat{
\varepsilon}_i \hat{
\varepsilon}_j  \tilde{N}_{i1}+
\frac{1}{N_2^2}
 \sum_{i=1}^N \hat{
\varepsilon}_i^2(1- \tilde{N}_{i1})+
\frac{1}{N_2^2}
\sum_{i=1}^N 
\sum_{j \neq i}
\hat{
\varepsilon}_i \hat{
\varepsilon}_j  (1-\tilde{N}_{i1})
(1-\tilde{N}_{j1})
\]
Again, whether or not the two moments are weighted does not impact the solution in case of uniqueness. 
This shows that the Two-moment estimator has the advantage and shortcoming that the information on the sign of individual deviations from their unconditional expected values is preserved in the second and forth term. The effect of a rare event combined with within-village correlation can cause a extreme deviation of villagers from their mean that in one direction. One would wish to employ the Two-moment estimator in the hope that it can mitigate such an effect, because agents in other villages might deviate in the opposite direction. One may advocate against the usage of the Two moment estimator if one suspects that it will even more attenuate the effect of such rare events.

\subsection{Consistency requirements}

Both estimates share the same consistency requirement. Convergence of both $\hat{Q}_{na}$ and $\hat{Q}_a$ to their respective limiting distributions require that 
\[ 
\frac{1}{N_m} \sum_{i=1}^N Y_{itm}I_{im} \overset{p}{\to}
p_0 \bar{r}_{m,tm}(q_0)
 \hspace{.25cm} \forall m=1,...,M
\]
A sufficient condition is that 
\[ \sum_{m=1}^M
\frac{1}{N_m} \sum_{i=1}^N Y_{itm}I_{im} \overset{p}{\to}
\sum_{m=1}^M 
w_m p_0 \bar{r}_{m,tm}(q_0)=
E\Bigg(\sum_{m=1}^M
\frac{1}{N_m} \sum_{i=1}^N Y_{itm}I_{im} \Bigg)
\] or equivalently 
\[ \sum_{m=1}^M
\frac{1}{N_m} \sum_{i=1}^N (Y_{itm}
- p_0 \bar{r}_{m,tm}(q_0))I_{im}
\overset{p}{\to}
0
\]
which warrants the applicability of a Law of Large Numbers. \\
\noindent Showing that the sum converges to its expected value despite correlation among outcomes requires showing that its variance shrinks to zero as $N$ goes to infinity.
A sufficient condition for this is that \[ Var(\frac{1}{N} \sum_{i=1}^N \hat{g}_i) \overset{p}{\to}
0
\]
Define 
\[ \bar{\sigma}^2=
\frac{1}{N_m} \sum_{i=1}^N
\hat{g}_i^2
\] as the sample Variance moment conditions and 
\[
\bar{\rho}_{v}=
\frac{1}{N_{2,v}(N_{2,v}-1)}
\sum_{i \in v}
\sum_{j \in v, i\neq j}
\hat{g}_i \hat{g}_j
(1-\tilde{N}_{i1})(1-\tilde{N}_{j1})
\] the observed correlation between  agents in village $v$. Note that correlation within the villages only concerns non-IPs, hence the division by $N_{2,v}(N_{2,v}-1)$ which is the largest possible pair of correlated agents in the village and the multiplication by one minus the IP-index. Then 
\[Var(\frac{1}{N} \sum_{i=1}^N \hat{g}_i)=
\frac{1}{N^2} N \bar{\sigma}^2 +
\frac{1}{N^2} \sum_{i=1}^N g_i
\sum_{j \neq i} g_j 
(1-\tilde{N}_{i1})(1-\tilde{N}_{j1})=
\]
\[ \frac{1}{N} 
\Big( \bar{\sigma}^2 +
\frac{\sum_v N_{2,v}(N_{2,v}-1)
\bar{\rho}_v}{N} \Big)
\]
which results from the fact that any correlation in individual outcomes (thus moment conditions) is correlation within the village. Consistency thus requires that
\[  \frac{1}{N} 
\Big( \bar{\sigma}^2 +
\frac{\sum_v N_{2,v}(N_{2,v}-1)
\bar{\rho}_v}{N} \Big) 
\overset{p}{\to} 0 
\]
implying that the second term in brackets must converge to a constant or decrease as the sample size increases to infinity. Intuitively, this implies that new villages that are added to the sample are not larger or more densely connected than the preceding ones. 
This condition is also sufficient to guarantee convergence of the variances for both estimators, albeit potentially at different rates. 

\subsection{Variance estimation and testing}

\citet{asycluster} derive conditions under which the Central Limit Theorem (CLT) can be applied to the sample mean in presence of clusters, e.g.\ whenever there are groups of observations that are correlated among themselves, but uncorrelated with the observations from other groups. Because both moment-based estimators are a function of the mean of the individual-specific moment conditions, hence they are a sample mean, this theory can be applied for the case at hand. Because individual-specific sample moment conditions are strictly bounded between minus one and one, thus under quite mild conditions, the sample moment covariance is a consistent estimator for the population moment covariance and the estimators are asymptotically normally distributed. The main insight here is that it is not necessary to know the limit of the sample covariance matrix as long as one is certain that it does  converge to that limit. This is sufficient for standard testing purposes for which neither heterogeneity nor correlation between moment conditions appears to be problematic. Nonetheless, as always with testing based on asymptotic theory, size distortions in small samples can be relevant and problematic. 

\section{Monte Carlo Study}

% \subsection{Study Setup}

Twelve real village networks are used for the simulation exercise. Following the enumeration of villages as done by the authors that accumulated the data, these are villages 1, 2, 4, 12, 23, 25, 31, 32, 45, 51, 57 and 73. The villages are densely connected. As a result, a substantial fraction of agents is directly linked to IPs and as such, they largely dominate the estimator. 
%Since the novelty of the moment-based estimator is the ability to extract information of later time periods, a comparison to the previously employed T2-estimator (which uses only the first information exchange for estimation) is only interesting if there is actually information to be gained when the time horizon is increased. Therefore, 
I limit the number of direct IP-neighbours by reducing the number of IPs to half of its original size. Accordingly, for each simulation run, the required number of IPs was randomly drawn from the observed set of IPs.
%I compare the T2 estimator with the non-aggregated moment-based estimator and the two-moment GMM estimator. 
Using the village network, the IP vector and the respective parameter combination, the outcome data was simulated. 
This way, 96 samples were generated and for each sample, the three estimation procedures were applied. 
The choice of the number of villages and simulation runs was guided by computational aspects: disposing over machines with 48 cores, three sample of villages could conveniently run in parallel. This way, 24 jobs in total could generate the desired estimates. 
To be able to run villages in parallel a common seed was first used to generate the village specific seeds. For the data simulation, this common seed equaled the simulation run (e.g. ranging from 1 to 96), for drawing the IPs, the common seed equaled the simulation run plus one (e.g.\ ranging from 2 to 97). 
Using the common seed, I generated twelve village seeds and picked the respective ones. Thereafter, the sample objective function is established by aggregating the village objective functions and the peak is identified by grid search. \\
\noindent 
%While the trimming estimate is harder to compute when there are more Ips, this is not in general true for the GMM estimator. When the number of IPs is very low, increasing this number indeed results in longer run-times as more agents are linked to the information to start with. Given the small average path length in the observed networks however, once the number of IPs has reached a slightly higher level, additional IPs decrease the computation time. 
Computational speed is decreasing in the number of IPs for the GMM estimator. This is intuitive: remembering that for direct and indirect neighbours, the reception probabilities are very easy to compute, a larger number of IPs directly translate into a faster evaluation of the objective function. In particular, for any agent three links away from the information, it is first necessary to check whether they fulfill the requirements of theorem 1 or theorem 2 in order to determine whether or not they can be used. It is these checks that take up a long time hence speed is slowest when many villagers are three links away from the IPs. \\ \noindent
Three parameter configurations were analyzed: $p_0=0.1, q_0=0.1, p_0=0.1, q_0=0.9$ and $p_0=0.5, q_0=0.5$. 
Beneath I plot the 96 estimates (blue) together with the mean estimate (red) and tabulate the bias as a percentage of the correct parameter value and the observed standard deviation.

\begin{figure}[H]
     \centering
       \caption{case 1: $p=0.1, q=0.1$}
  %   \begin{subfigure}[b]{0.49\textwidth}      \centering         \includegraphics[width=\textwidth]{0101/t2.jpg}
        % \caption{T2-estimates} \end{subfigure} 
     %\hfill
     \begin{subfigure}[b]{0.49\textwidth}
         \centering
         \includegraphics[width=\textwidth]{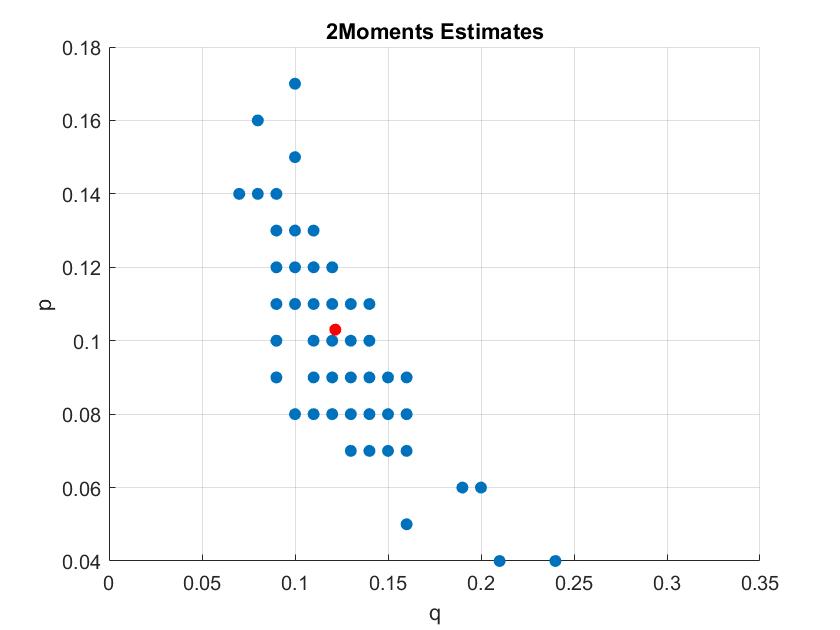}
       %  \caption{2-Moment Estimates}
     \end{subfigure} 
  %   \hfill
     \begin{subfigure}[b]{0.5\textwidth}
         \centering
         \includegraphics[width=\textwidth]{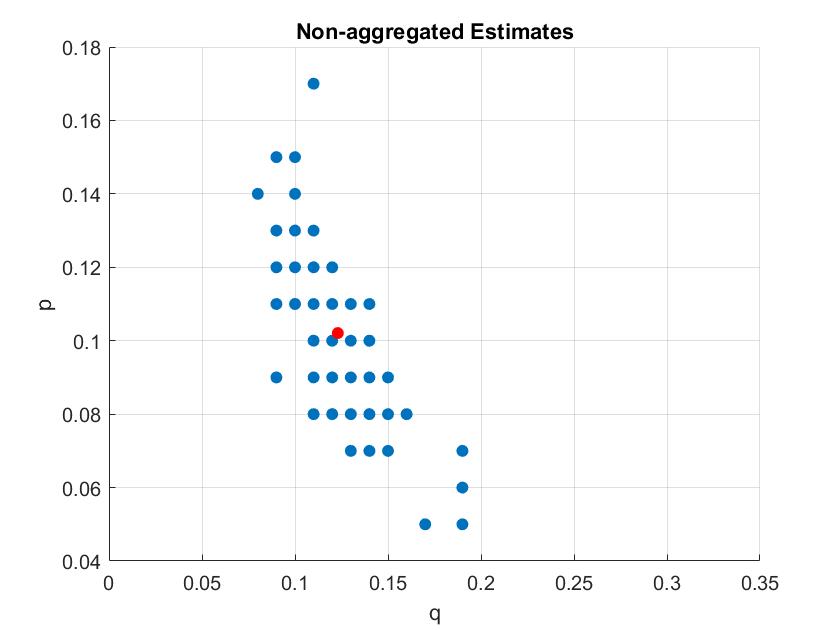}
      %   \caption{Non-aggregated Estimates}
     \end{subfigure}
\end{figure}

\begin{table}[ht]
\begin{tabular}{l|l|l|l|l|l|l}
 & \multicolumn{6}{c}{p=0.1, q=0.1}                         \\ \hline
& mean p & \% bias & std. dev.   & mean q & \% bias & std. dev.  \\ \hline
%T2   & 0.1031 & 3.1     & 0.0265 & 0.11   & 10      & 0.0495 \\
Non aggregated estimator        & 0.1021 & 2.1     & 0.0237 & 0.123  & 23      & 0.0229 \\
2 moments estimator            & 0.103  & 3       & 0.0259 & 0.1217 & 21.7    & 0.028 
\end{tabular}
\end{table}

\begin{figure}[H]
     \centering
       \caption{case 2: $p=0.1, q=0.9$}
  %   \begin{subfigure}[b]{0.49\textwidth}      \centering         \includegraphics[width=\textwidth]{0109/t2.jpg}
        % \caption{T2-estimates} \end{subfigure} 
   %  \hfill
     \begin{subfigure}[b]{0.49\textwidth}
         \centering
         \includegraphics[width=\textwidth]{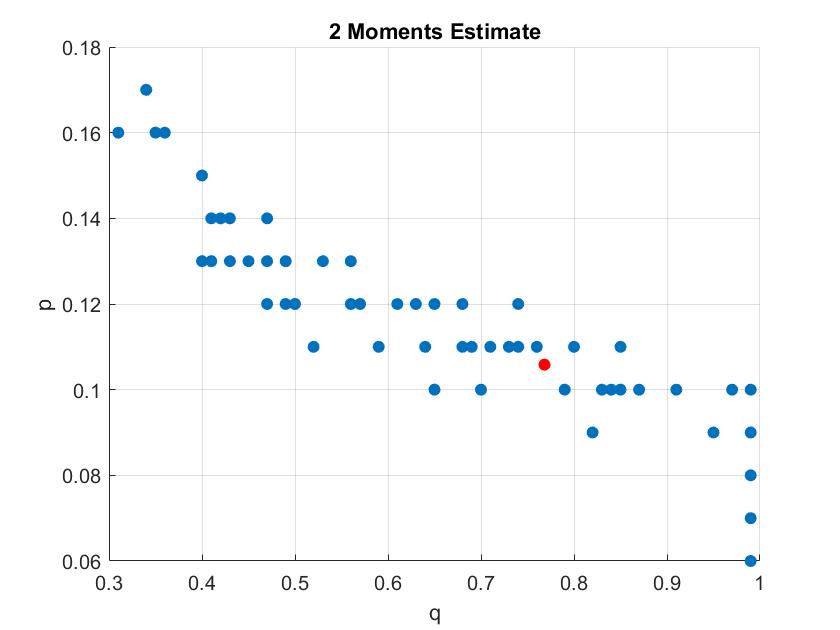}
       %  \caption{2-Moment Estimates}
     \end{subfigure} 
  %   \hfill
     \begin{subfigure}[b]{0.49\textwidth}
         \centering
         \includegraphics[width=\textwidth]{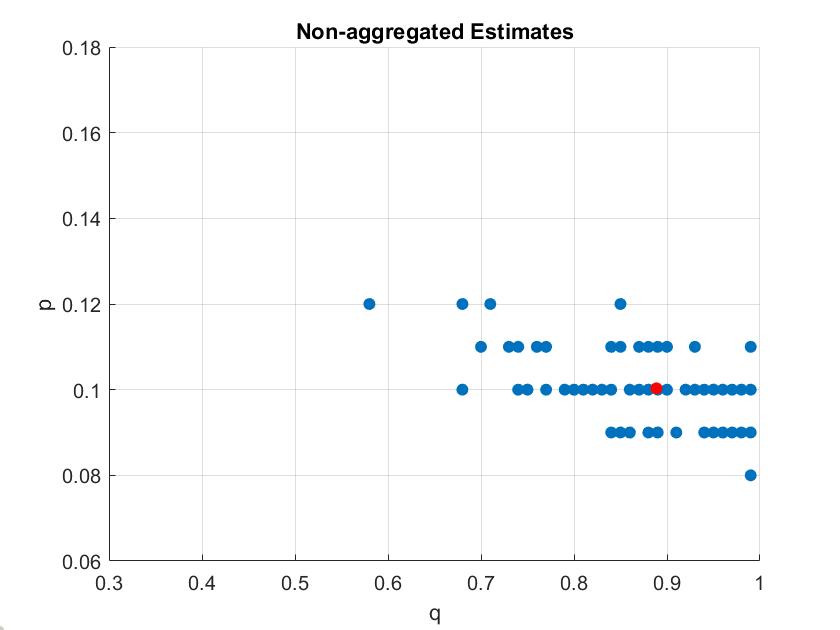}
      %   \caption{Non-aggregated Estimates}
     \end{subfigure}
\end{figure}

\begin{table}[h!]
\begin{tabular}{l|l|l|l|l|l|l}
                      & \multicolumn{6}{c}{p=0.1,q=0.9}                           \\ \hline
                      & mean p & \% bias & std. dev.   & mean q & \% bias  & std. dev.  \\ \hline
%T2                    & 0.1057 & 5.7     & 0.0148 & 0.8432 & 6.311111 & 0.1418 \\
Non aggregated estimator        & 0.1003 & 0.3     & 0.0077 & 0.8886 & 1.266667 & 0.0901 \\
2 moments estimator            & 0.1058 & 5.8     & 0.0222 & 0.7682 & 14.64444 & 0.2249
\end{tabular}
\end{table}

\begin{figure}[H]
     \centering
       \caption{case 3: $p=0.5, q=0.5$}
  %   \begin{subfigure}[b]{0.49\textwidth}      \centering         \includegraphics[width=\textwidth]{0505/t2.jpg}
        % \caption{T2-estimates}\end{subfigure} 
     \hfill
     \begin{subfigure}[b]{0.49\textwidth}
         \centering
         \includegraphics[width=\textwidth]{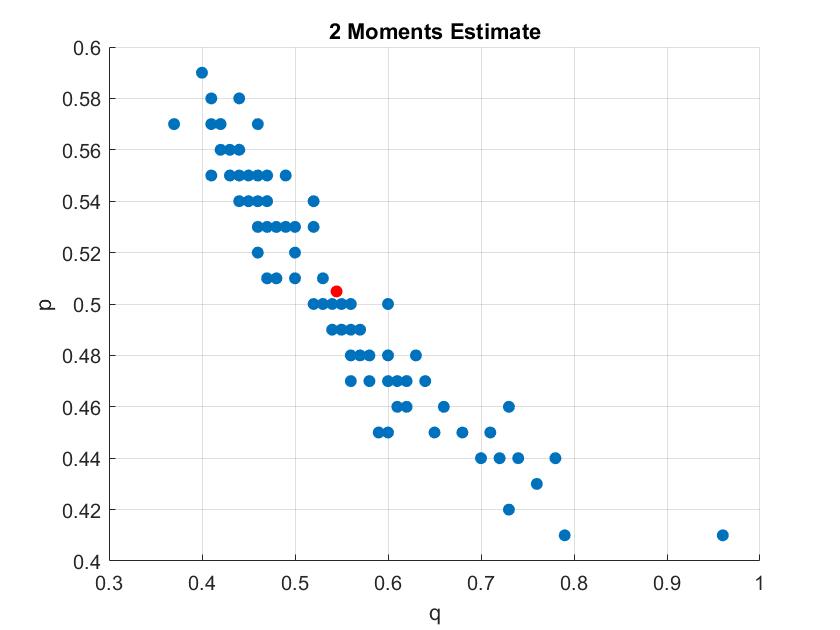}
       %  \caption{2-Moment Estimates}
     \end{subfigure} 
  %   \hfill
     \begin{subfigure}[b]{0.49\textwidth}
         \centering
         \includegraphics[width=\textwidth]{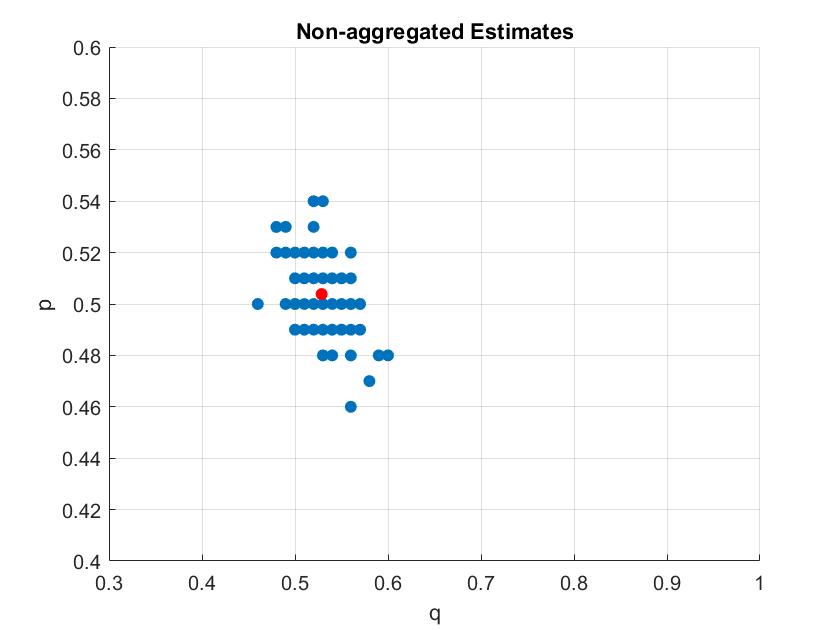}
      %   \caption{Non-aggregated Estimates}
     \end{subfigure}
\end{figure}

\begin{table}[ht]
\begin{tabular}{l|l|l|l|l|l|l}
                      & \multicolumn{6}{c}{p=0.5, q=0.5}                          \\ \hline
                      & mean p & \% bias & std. dev.   & mean q & \% bias & std. dev.  \\ \hline
%T2                    & 0.5071 & 1.42    & 0.0385 & 0.4994 & 0.12    & 0.0551 \\
Non aggregated estimator        & 0.5039 & 0.78    & 0.0156 & 0.5286 & 5.72    & 0.0263 \\
2 moments estimator            & 0.5049 & 0.98    & 0.0441 & 0.5447 & 8.94    & 0.1065
\end{tabular}
\end{table}

\noindent
Visibly, a substantial reduction in the spread of the estimates can be achieved with the non-aggregated estimator. In line with this, standard deviations are lowest for this estimator. In case 1, the non-aggregated estimator for $q$ is biased, but the plot reveals that this is more the effect of some outliers, rather than a general property. Interestingly, this is the case for all three parameter configurations, but most pronounced in case 2. This shows that first some gains can be made when later time periods are considered and that within-village correlations increase the estimator variance in finite samples, in line with the derivations above. It has to be kept in mind that the superiority of the non-aggregated estimator comes with the additional advantage of the possibility of parallelization over villages which substantially increases speed.  

\section{Application}

For the application, all 37 villages were used. A previous estimate had generated $\hat{p}=0.16, \hat{q}=0.79$. Above are surface plots for both estimation techniques. Unsurprisingly, the Non-aggregated estimator is very close to the previous estimator. As mentioned above, the number of IPs is large and as a consequence, many villagers are directly connected to them and little can be gained by increasing the time horizon. The bad performance of the Two-Moment Estimator is more surprising and highlighting once more that within-village correlation is important, advocating that aggregating individual moments in a GMM-style estimation can be problematic in the ``latent-diffusion-observed-adoption" model.       

\begin{figure}
\caption{Real village estimation}
    \centering
    \includegraphics[scale=0.2]{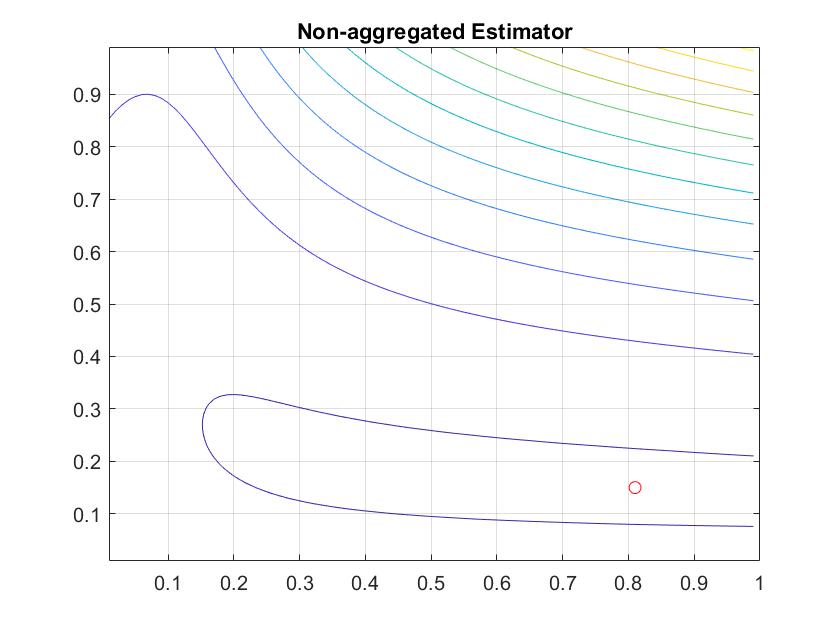}
    \includegraphics[scale=0.2]{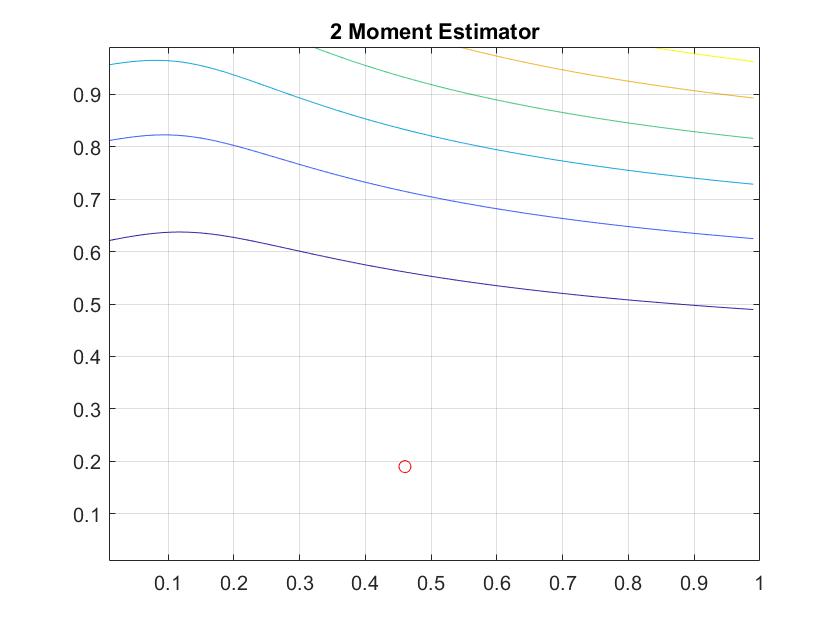}
\end{figure}

\section{Conclusion}

Applying moment-based estimation to the network diffusion model with unobserved information propagation is potentially beneficial, yet challenging. The individuals' information reception probabilities are in general hard to evaluate. The problem can be tackled by using only one observation per individual, namely the one of the period that immediately follows the first information exchange in which the agent could have been informed. Here, some short-hand formulas exist, making the computations tractable. 
This leads to individual-specific moment conditions that can be used to set up an objective function.  
\\ \noindent
As mean and variances of the outcomes are link-portfolio specific and correlation across individual means arises through the village network, aggregating non-IP individuals into moment conditions can result in small sample bias. 
Under regularity conditions outlined in \cite{asycluster}, the researcher is able to obtain an unbiased estimate of the covariance of the moment conditions and hence the parameters. Alternatively, when villages are sufficiently small, the researcher may choose to compute each village covariance matrix which is fully specified in terms of the model parameters and the network. Finally, when villages become very numerous, the covariance matrix of the moment conditions is sparse and the researcher may choose to neglect its off diagonal non-zero entries. It remains to be investigated which of these proceedings has the best performance.
\\ \noindent 
The Monte Carlo results are in line with the theoretical derivations, providing evidence that aggregating individual moments - as in a usual GMM framework - can lead to a large spread of the estimates. \\
\noindent Given the promising results from using later time periods, further research in finding shorthand formulas that enable a further extension of the modeled time horizon has the potential to be highly beneficial. 

\section{Appendix - Proofs of Theorems 1 and 2}

\subsection{Proof of Theorem 1}

\begin{proof}
For the seek of clarity, I assume that all IP neighbours have the same number of links to IPs, although the proof can also be established in general. 
Let 
\[ E(r_{it}(q))=\sum_{S_{1:(t-1)}\in \mathbf{S}} E(S_{it}|S_{1:(t-1)},G,q)P(S_{(t-1)},G,q)\]
the information reception rate achieved by integrating out the latent variable and let $\omega_i$ be the receiver's degree and let $\Bar{\omega}$ be the senders' degree, same for all senders by assumption (i.e. $\omega_j=\Bar{\omega} \forall j=1,..,\omega_i$). I need to account for all information scenarios that imply that $i$ has a nonzero probability to be informed. First, she must have some informed neighbours. Given that she has $\omega_i$ neighbours in total, the number of informed neighbours she has, which I denote $x$, can vary from zero to $\omega_i$. Each time, there are various (precisely $ \binom{\omega_i}{x}$) 
possibilities to draw $x$ out of $\omega_i$ individuals. For simplicity I have assumed that the $\omega_i$ individuals on the path between $i$ and the IP are homogeneous in the sense that each has herself degree $\bar{\omega}$. Therefore, each of them has, by formula \eqref{rr}, probability $(1-(1-q)^{\bar{\omega}})$ to be informed in the first exchange and probability $(1-q)^{\bar{\omega}}$ to enter the second period uninformed. Then   
$\big(1-(1-q)^{\Bar{\omega}}\big)^x
\big((1-q)^{\Bar{\omega}}\big)^{\omega^i-x}$ is the probability that $x$ out of the $\omega_i$ friends are informed, while the remaining $\omega_i-x$ stay uninformed.
So far I have established the probability that the second information exchange starts with $x$ out of $\omega_i$ friends of $i$ being informed. I still need to evaluate what in this case will be $i's$ probability to receive the information.
$(1-(1-q)^x)$ is $i's$ chance to be informed, given that $x$ out of her $\omega_i$ friends are informed. Letting $x$ vary from 0 to $\omega_i$ and summing up over all possibilities, this leads to
\[ E(r_{it}(q,S_{1:t-1}))=
\sum_{x=0}^{\omega_i}
\binom{\omega_i}{x}
\big(1-(1-q)^x\big)
\big(1-(1-q)^{\Bar{\omega}}\big)^x
\big((1-q)^{\Bar{\omega}}\big)^{\omega_i-x}
\]
To recapitulate: the binomial coefficient gives the number  possibilities how $x$ informed neighbours can be drawn out of the $\omega_i$ available ones, $\big(1-(1-q)^x\big)$ gives $i's$ probability to receive the information given she has $x$ informed neighbours and  $\big(1-(1-q)^{\Bar{\omega}}\big)^x
\big((1-q)^{\Bar{\omega}}\big)^{\omega_i-x} $ is the probability that indeed $x$ of her neighbours are informed and $\omega_i-x$ stay uninformed. 
The total probability mass is then obtained by summing over all possibilities for $x$. Evidently, if $x=0$, then $r_{it}=0$. 
\[ r_{it}=
\sum_{x=0}^{\omega_i}
\binom{\omega_i}{x}
\big(1-(1-q)^{\Bar{\omega}})\big)^x
\big((1-q)^{\Bar{\omega}}\big)^{\omega_i-x}
\]
\[ -
\sum_{x=0}^{\omega_i}
\binom{\omega^i}{x}
\big((1-q)^x)\big)
\big(1-(1-q)^{\Bar{\omega}})\big)^x
\big((1-q)^{\Bar{\omega}}\big)^{\omega_i-x}
\] Using the formula for the binomial coefficient 
\[= \Big(1-(1-q)^{\Bar{\omega}}+(1-q)^{\Bar{\omega}}\Big)^{\omega_i}
- \sum_{x=0}^{\omega_i}
\binom{\omega_i}{x}
\big((1-q)(1-(1-q)^{\Bar{\omega}})\big)^x
\big((1-q)^{\Bar{\omega}}\big)^{\omega_i-x}
\]
\[ =1- 
\Big(
(1-q)(1-(1-q)^{\Bar{\omega}}+(1-q)^{\Bar{\omega}}\Big)^{\omega_i}
\]
\[ =1- 
\Big(1 - (1-q)^{\Bar{\omega}}
-q + q(1-q)^{\Bar{\omega}}+(1-q)^{\Bar{\omega}}\Big)^{\omega^i}
\]
\[=1-
\Big(1 
-q + q(1-q)^{\Bar{\omega}}\Big)^{\omega_i}
\]
\[= 1-(1-q(\underbrace{1-(1-q)^{\Bar{\omega}}}_{
\Bar{r}_{j(t-1)}}
)^{\omega_i} 
= \Bar{r}_{it}
\]
\end{proof}

\subsection{Proof of Theorem 2}

\begin{proof}
For clarity I assume that the IP neighbours linked to $i$  are homogeneous, each having $J$ links to the final agent and the same number of links to IPs and that there are in total $\omega_i$ such agents that are linked to the IPs and indirectly linked to $i$. By the assumptions of theorem 2, none of the agents on the (2 link) path between the IP-neighbours and the final agent is frequented twice and as such, these paths can be treated as independent. 

\[ E(r_{i3}(q)=
\sum_{x=0}^{\omega_i}
\binom{\omega_i}{x}
(1-(1-q^2)^{Jx})
 (\Bar{r}_{k1})^x
\big(1-\Bar{r}_{k1} \big)^{\omega_i-x}
\]
The binomial coefficient again gives the number of possibilities to draw $x$ out of $\omega_i$ IP-neighbours to be informed. 
The probability that the final agent receives the information through any one particular path is $q^2$ and paths can be treated independently. 
If $x$ IP neighbours are informed and each has $J$ paths to the final agent, then the total number of paths is $Jx$ and consequently $i's$ probability to receive the information through any path is $(1-(1-q^2)^{Jx})$ e.g. one minus the probability that she does not receive the information. 
Further $ (\Bar{r}^k_1)^x
\big(1-\Bar{r}^k_1 \big)^{\omega_i-x}$ gives the probability that $x$ out of the $\omega_i$ IP neighbours linked to $i$ are informed, while the remaining ones stay uninformed. 
\[ =
\sum_{x=0}^{\omega_i}
\binom{\omega_i}{x}
 (\Bar{r}_{k1})^x
\big(1-\Bar{r}_{k1}\big)^{\omega_i-x}
\]

\[ -
\sum_{x=0}^{\omega^i}
\binom{\omega^i}{x}
((1-q^2)^{Jx})
 (\Bar{r}_{k1})^x
\big(1-\Bar{r}_{k1} \big)^{\omega_i-x}
\]

\[ =1 -
\sum_{x=0}^{\omega_i}
\binom{\omega_i}{x}
((1-q^2)^{J}
 \Bar{r}_{k1})^x
\big(1-\Bar{r}_{k1} \big)^{\omega_i-x}
\]

\[1-\big( (1-q^2)^J\Bar{r}_{k1}+1-\Bar{r}_{k1} \big)^{\omega_i}
\]

\[ 
1- \big(1-\Bar{r}_{k1}(1-(1-q^2)^J) \big)^{\omega_i}
= \Bar{r}_{i3}
\]

\end{proof}

\subsection{First Order Conditions (Non-aggregated Estimator)}

\[ 
\frac{\partial(-Q)}{\partial p }=
2 \sum_{t=2}^T \sum_{i \in \tilde{P}}
r_{i(t-1)}-
2 \sum_{t=2}^T \sum_{i \in \tilde{N}}
(r_{i(t-1)})^2p +2 \# IPP 
- 2 \# IP p -  =0
\]

\begin{equation}
\label{FOCp}
    p=\frac{
    \sum_{t=2}^T \sum_{i \in \tilde{P}}
(r_{i(t-1)}) + \#IPP
    }{\sum_{t=2}^T \sum_{i \in \tilde{N}}
(r_{i(t-1)})^2 + \#IP}
\end{equation}

From \ref{FOCp} it is apparent that the optimal $p$ exceeds one if 
\[ \#IPP - \#IP <
\sum_{t=2}^T \sum_{i \in \tilde{N}}
(r_{i(t-1)})^2-\sum_{t=2}^T \sum_{i \in \tilde{P}}
(r_{i(t-1)})
\]
\[ \#IPP - \#IP <
\sum_{t=2}^T \sum_{
\substack{i \in \tilde{N} \\ i \notin \tilde{P}}
}
(r_{i(t-1)})^2
+ \sum_{t=2}^T \sum_{i \in \tilde{P}}
\big(
(r_{i(t-1)})^2-(r_{i(t-1)})\big)
\]
The left hand side (LHS) of this inequality is always non-positive. The first term on the right hand side (RHS) is positive and increasing in the number of paths leading to $i$ (her "information degree"). The second term on the RHS is negative but  converges to zero for large enough information degrees. \\ \noindent
This shows that the number of links (and thus the signal reception rates) among non-ips in general and non-ip participants in particular must be sufficiently large to prevent a corner solution.  
\[ \frac{\partial (-Q)}{\partial q}=
2 \sum_{t=2}^T \sum_{i \in \tilde{P}}
p \frac{\partial r_{i(t-1)}}{\partial q}
- 2 \sum_{t=2}^T \sum_{i \in \tilde{N}} p^2 
r_{i(t-1)} \frac{\partial r_{i(t-1)}}{\partial q} =0
\]
\begin{equation}
    \label{FOCq}
    p= \frac{\sum_{t=2}^T \sum_{i \in \tilde{P}}
    \frac{\partial r_{i(t-1)}}{\partial q}
    }{
    \sum_{t=2}^T \sum_{i \in \tilde{N}}
    r_{i(t-1)}
    \frac{\partial r_{i(t-1)}}{\partial q}
    }
\end{equation}
Intuitively, the RHS tends to zero when $q$ is large and there are few participants that possess many links to the information. \\
\noindent Combining the FOCs leads to
\begin{equation}
    \label{FOCs}
    \frac{
    \sum_{t=2}^T \sum_{i \in \tilde{P}}
(r_{i(t-1)}) + \#IPP
    }{\sum_{t=2}^T \sum_{i \in \tilde{N}}
(r_{i(t-1)})^2 + \#IP}=\frac{\sum_{t=2}^T \sum_{i \in \tilde{P}}
    \frac{\partial r_{i(t-1)}}{\partial q}
    }{
    \sum_{t=2}^T \sum_{i \in \tilde{N}}
    r_{i(t-1)}
    \frac{\partial r_{i(t-1)}}{\partial q}
    }
\end{equation}
\ref{FOCs} pins down $q$ as a function of the number of IPs, the number of IP participants as well as the agents in $\tilde{N}_2,..,\tilde{N}_T,\tilde{P}_2,..,\tilde{P}_T, $ and their respective link portfolios. 
Naturally, some variation in the outcomes for non-ips is required, else the system of FOCs has no solution.

\subsection{Convexity of the Objective Function (Non-aggregated Estimator)}

The objective function $-Q$ is convex if and only if the Hessian is positive semi definite. 
For the Hessian to be positive semi definite, the determinant is required to be positive and which in turn requires that 
\[ Det(\mathcal{H})= 
\frac{\partial^2 Q}{\partial^2 p}
\frac{\partial^2 Q}{\partial^2 q}-
\Big( \frac{\partial^2 Q}{\partial p \partial q}
\Big)^2
>0
\]
Since $Det(-Q)=Det(Q)$. 
\[ Det(\mathcal{H})=\]
\[\underbrace{\left( 2 \sum_{t=2}^T \sum_{i \in \tilde{N}_t }
(r_{i(t-1)})^2 + \# IP \right)}_{\frac{\partial^2 Q}{\partial^2 p}}
\underbrace{
\left( 2 \sum_{t=2}^T \sum_{i \in \tilde{N}_t } p^2
\Bigg( \Bigg[ \frac{\partial r_{i(t-1)}}{\partial q}\Bigg]^2 +r_{i(t-1)} \frac{\partial^2r }{\partial^2 q} \Bigg) 
-2 \sum_{t=2}^T \sum_{i \in \tilde{P}_t}
p \frac{\partial^2 r_{i(t-1)}}{\partial^2 q} \right)
}_{\frac{\partial^2 Q}{\partial^2 q}}
\]
\[ - \underbrace{
 \left( 4 \sum_{t=2}^T \sum_{i \in \tilde{N}_t }
p r_{i(t-1)} \frac{\partial r_{i(t-1)}}{\partial q} - 2 \sum_{t=2}^T \sum_{i \in \tilde{P}_t }
\frac{\partial r_{i(t-1)}}{\partial q}
\right)^2
}_{\Big( \frac{\partial^2 Q}{\partial p \partial q}
\Big)^2
}
\]
A necessary condition is that 
\[ 
\left( 2 \sum_{t=2}^T \sum_{i \in \tilde{N}_t } p^2
\Bigg( \Bigg[ \frac{\partial r_{i(t-1)}}{\partial q}\Bigg]^2 +r_{i(t-1)} \frac{\partial^2r }{\partial^2 q} \Bigg) 
-2 \sum_{t=2}^T \sum_{i \in \tilde{P}_t}
p \frac{\partial^2 r_{i(t-1)}}{\partial^2 q} \right)>0
\]
\begin{equation}
    \label{detcon}
      \sum_{t=2}^T \sum_{i \in \tilde{N}_t } p
\Bigg( \Bigg[ \frac{\partial r_{i(t-1)}}{\partial q}\Bigg]^2 +r_{i(t-1)} \frac{\partial^2r }{\partial^2 q} \Bigg) 
>  \sum_{t=2}^T \sum_{i \in \tilde{P}_t}
\frac{\partial^2 r_{i(t-1)}}{\partial^2 q} 
\end{equation}
The first derivative of $r_{i(t-1)}$ with respect to $q$ depends on $w^i$, the number of paths to the information that individual $i$ possesses. It is increasing in $w^i$ for small values of $q$, decreasing in $w^i$ in the large-$q$ area and first increasing, then decreasing for intermediate values of $q$. In any case, the slope decreases almost everywhere and hence the second derivative is usually  negative, implying that the RHS of \ref{detcon} is negative. The absolute value of the RHS increases in the number of non-ip participants and in the number of ip connections they possess. As a consequence, it helps fulfilling the convexity condition if non ip participants are numerous and densely connected with the ips.  \\
\noindent On the left-hand-side (LHS) the term 
\[ 
\Bigg[ \frac{\partial r_{i(t-1)}}{\partial q}\Bigg]^2 +r_{i(t-1)} \frac{\partial^2r }{\partial^2 q} 
\]
is positive for small and negative for larger values of $w^i$ and the intersection with the horizontal axis occurs the earlier the higher $q$. The function is first positive, crosses the axis and becomes negative and then exhibits another turning point such that it finally converges to zero for very large values of $w^i$. As a consequence, heterogeneity in $w^i$ helps fulfilling the convexity condition as the function is positive for small $w^i$ and close to zero for large $w^i$.
\\ \noindent
When all agents exhibit intermediate values of $w^i$  and/or participation rates are low, on the other hand, the convexity condition may fail to hold.
 Observe that a large number of non-ip participants also decreases the cross partial derivative, and as such also that the determinant is positive. 
The convexity condition is harder to fulfill in the high-$q$ area. \\
\noindent Given a particular $p,q$ point, a certain number of agents that contribute to the objective function $\tilde{N}$ and a particular degree distribution among them, equation \ref{detcon} can be used to determine the minimal level of non-ip-participation that guarantees convexity of the objective function. 
% NB the derivates for all waves are similar in shape 
\\
\noindent
Convexity of the objective function can also be investigated by means of a Taylor series expansion, the aim being to demonstrate that higher order terms vanish. \\
\noindent The third order terms are 
\begin{equation}
    \label{T1}
    \frac{1}{2}
\frac{\partial^3 Q}{\partial^2 p \partial q}
= 2 \sum_{t=2}^T \sum_{i \in \tilde{N}_t }
r_{i(t-1)} \frac{\partial r_{i(t-1)}}{\partial q}
\end{equation}
\begin{equation}
    \label{T2}
    \frac{1}{2}
\frac{\partial^3 Q}{\partial p \partial^2 q} =
2 p \sum_{t=2}^T \sum_{i \in \tilde{N}_t }
\Bigg( \Bigg[ \frac{\partial r^i_{(t-1)}}{\partial q}\Bigg]^2 +r_{i(t-1)} \frac{\partial^2r }{\partial^2 q} \Bigg) 
-  \sum_{t=2}^T \sum_{i \in \tilde{P}_t}
\frac{\partial^2 r_{i(t-1)}}{\partial^2 q}
\end{equation}
\[  \frac{1}{6}
\frac{\partial^3 Q}{\partial^3  q}= \]
\begin{equation}
    \label{T3}
\frac{1}{3} p^2 \sum_{t=2}^T \sum_{i \in \tilde{N}_t }
\Bigg( 3 \frac{\partial r_{i(t-1)}}{\partial
 q}
 \frac{\partial^2 r_{i(t-1)}}{\partial^2
 q}
 +
 r_{i(t-1)}
 \frac{\partial^3 r_{i(t-1)}}{\partial^3
 q} \Bigg)
- \frac{1}{3 }
\sum_{t=2}^T \sum_{i \in \tilde{P}_t }
p \frac{\partial^3 r_{i(t-1)}}{\partial^3 q}
\end{equation}
\begin{equation}
    \label{T4}\frac{1}{6}
\frac{\partial^3 Q}{\partial^3 p }=0
\end{equation}
First, observe that the shape of 
$r_{i(t-1)}
\frac{\partial r_{i(t-1)}}{\partial q}$
and $\frac{\partial r_{i(t-1)}}{\partial q}\frac{\partial^2 r_{i(t-1)}}{\partial^2 q}$ are reflections of one another about the horizontal axis and hence mitigate one another. This is because the first derivative is increasing and  while the second derivative is decreasing. Further,  if $Det(\mathcal{H})>0$, then \ref{T2} is positive. The third derivative is positive again. The forth term of \ref{T3} can mitigate the third term of \ref{T3} and \ref{T2} only if non-ip participants are sufficiently numerous.

\section{Bibliography}
\nocite{*}
\bibliographystyle{chicago}
\bibliography{my}

\end{document}